\documentclass[epsfig,12pt,onecolumn]{article}
\usepackage{amsfonts}
\usepackage{amssymb}
\usepackage{multicol}
\usepackage{graphicx}
\usepackage{float}
\usepackage{caption}
\usepackage{xcolor}
\usepackage[utf8]{inputenc}
\usepackage{amsmath}

\makeatletter \@addtoreset{equation}{section}
\renewcommand{\theequation}{\thesection.\arabic{equation}}
\def \be{\begin{equation}}
\def \ee{\end{equation}}
\def \bea{\begin{eqnarray}}
\def \eea{\end{eqnarray}}

\newcommand{\nc}{\newcommand}
\nc{\al}{\alpha} \nc{\bib}{\bibitem} \nc{\la}{\lambda}
\nc{\C}{\mbox{\hspace{1.24mm}\rule{0.2mm}{2.5mm}\hspace{-2.7mm} C}}
\nc{\R}{\mbox{\hspace{.04mm}\rule{0.2mm}{2.8mm}\hspace{-1.5mm} R}}
\input{tcilatex}
\begin{document}

\title{\rightline{\mbox{\small {LPHE-MS-Sept-24-revised}} \vspace
{-0.2 cm}} \textbf{Topological 4D gravity and }\\
\textbf{gravitational defects }}
\author{Y. Boujakhrout \footnote{boujakhroutyoussra@gmail.com}, R. Sammani, E.H Saidi \\
%EndAName
{\small 1. LPHE-MS, Science Faculty}, {\small Mohammed V University in
Rabat, Morocco}\\
{\small 2. Centre of Physics and Mathematics, CPM- Morocco}}
\maketitle

\begin{abstract}
Using the Chern-Simons formulation of AdS$_{3}$ gravity as well as the
Costello-Witten-Yamazaki (CWY) theory for quantum integrability, we
construct a novel topological 4D gravity given by Eq(\ref{4sd}) with
observables based on gravitational gauge field holonomies. The field action $%
\mathcal{S}_{4D}^{grav}$ of this gravity has a gauge symmetry $SL(2,\mathbb{C%
})$ and reads also as the difference $\mathcal{S}_{4D}^{CWY_{L}}-\mathcal{S}%
_{4D}^{CWY_{R}}$\ with 4D Chern-Simons field actions $\mathcal{S}%
_{4D}^{CWY_{L/R}}$ given by left/right CWY theory Eq(\ref{LCWY}). We also
use this 4D gravity derivation to build observables describing gravitational
topological defects and their interactions.\textrm{\ }We\textrm{\ conclude
our study with few comments regarding quantum integrability and the
extension of AdS}$_{3}$\textrm{/CFT}$_{2}$\textrm{\ correspondence with
regard to the obtained topological 4D gravity.}

\textbf{Keywords}: \emph{AdS}$_{3}$\emph{\ gravity and CS formulation, AdS}$%
_{3}$\emph{/CFT}$_{2}$\emph{, Line defects in CS and 4D gravity, Integrable
spin chains and brane realisations in strings.}
\end{abstract}

%\tableofcontents

\section{Introduction}

\qquad Following \textrm{\cite{1A}}, the 3D Chern-Simons theory with
hermitian gauge field action $\mathcal{S}_{3D}^{{\small CS}}\left[ A\right] $
has an unconventional 4D extension giving a powerful QFT framework to study
quantum integrability \textrm{\cite{2A}-\cite{2AC}}. This is an exotic 4D
extension of the usual 3D Chern-Simons (CS) modeling to which we refer below
\textrm{to} as the Costello-Witten-Yamazaki theory. It lives in a 4D space $%
\boldsymbol{M}_{4D}$ fibered like $\Sigma _{2}\times \tilde{\Sigma}_{2}$;
that is the products of two 2D real surfaces thought of in this study as $%
\mathbb{R}^{2}\times \mathbb{CP}^{1}$ with complex projective line
isomorphic to the real 2-sphere $\mathbb{S}^{2}$ \textrm{\cite{1B}}. The
Costello-Witten-Yamazaki (CWY) theory is a 4D topological quantum field
theory with field action $\mathcal{S}_{4D}^{{\small CWY}}\left[ \mathbf{A}%
\right] $ leading to a flat 2-form gauge curvature \textbf{F}$_{2}=d\mathbf{A%
}+\mathbf{A}^{2}=0$. Its observables are given by line and surface defects
that have been shown to carry precious information on quantum integrability
\cite{PF1}-\cite{PF5}. Typical gauge invariant line defects are given by
Wilson and 't Hooft lines as well as their braiding \textrm{\cite{1BC},\cite%
{1BD}}. These line defects have been given interpretations in terms of Q-operators of integrable
quantum spin and superspin chains \textrm{\cite{1C}}; and were realised
in terms of intersecting branes in type II strings and M-theory \textrm{\cite%
{1D}-\cite{1DB}}. The main steps in the derivation of the CWY theory by
starting from the 3D Chern-Simons are given in section 3; we refer to these
steps as the Building ALgorithm (BAL); this BAL will be in our construction.
\ \ \

On the other hand, it is quite well known that 3D Anti-de Sitter (AdS$_{3}$)
gravity is intimately related with 3D Chern-Simons theory \textrm{\cite{1E}-%
\cite{2EAC}}. The field action $\mathcal{S}_{AdS_{3}}^{grav}\left[ \mathbf{%
e,\omega }\right] $ of the AdS$_{3}$ gravity is a functional of two 1-forms
given by the real dreibein $\mathbf{e}$\ and the real 3D spin connection $%
\mathbf{\omega }$. But it has been shown that $\mathbf{e}$ and$\ \mathbf{%
\omega }$ can be expressed in terms of left/right pair ($A_{L},A_{R}$) of
Chern-Simons gauge fields in 3D. In this CS description, the $\mathbf{\omega
}$ is given by the mean field ($A_{L}+A_{R}$)/2 and $\mathbf{e}$ by the
reduced field variable ($A_{L}-A_{R}$)/2. Concretely, it was found that the $%
\mathcal{S}_{AdS_{3}}^{grav}$ can be nicely formulated like the difference
of two Chern-Simons field actions as
\begin{equation}
\mathcal{S}_{AdS_{3}}^{grav}=\mathcal{S}_{3D}^{{\small CS}}\left[ A_{L}%
\right] -\mathcal{S}_{3D}^{{\small CS}}\left[ A_{R}\right]
\end{equation}%
where $A_{L}$ and $A_{R}$ are hermitian 1-form CS\ gauge potentials valued
in left and right Lie algebras; say $A_{L}$ valued in $sl(2,\mathbb{R})_{L}$
and $A_{R}$ valued in $sl(2,\mathbb{R})_{R}.$ Because of its rich
properties, the AdS$_{3}$ gravity has been the subject of increasing
interest; especially in connection with AdS$_{3}$/CFT$_{2}$ correspondence
\textrm{\cite{2AA,2EAB,Vero}, }BTZ black holes \textrm{\cite{3A} }and higher
spin 3D gravities\textrm{\ \cite{2AA,2EAB,4A}}. \ \ \

In this paper, we contribute to the study of AdS$_{3}$ gravity from the view
of quantum integrability and topological gravitational defects. For that, we
apply the method of CWY of integrable systems to the AdS$_{3}$ gravity with
Anti-de Sitter group $SO(2,2)$ while using its formulation in terms of the
CS gauge potentials $A_{L/R}$. In this way, one opens a window on
applications of integrable system methods obtained by CWY in QFT to 4D
topological gravity with line and surface defects. The outcome of this study
is twofold: \ \ \ \

First, the derivation of a novel 4D topological gravity living on the 4D
space $\mathbb{R}^{2}\times \mathbb{CP}^{1}$ with field action $\mathcal{S}%
_{4D}^{grav}$ that can be expressed in two equivalent ways: $\left( \mathbf{i%
}\right) $ as a functional of a 4D left $\mathbf{A}_{L}$ and a 4D right $%
\mathbf{A}_{R}$ Chern-Simons like fields (two CWY gauge fields $\mathbf{A}%
_{L}/\mathbf{A}_{R}$) as follows%
\begin{equation}
\mathcal{S}_{4D}^{grav}\left[ \mathbf{A}_{L}\mathbf{,A}_{R}\right] =\mathcal{%
S}_{4D}^{{\small CWY}}\left[ \mathbf{A}_{L}\right] -\mathcal{S}_{4D}^{%
{\small CWY}}\left[ \mathbf{A}_{R}\right]  \label{grav}
\end{equation}%
with $\mathcal{S}_{4D}^{{\small CWY}}\left[ \mathbf{A}_{L/R}\right] $ given
by eqs(\ref{4dg}-\ref{R}); or $\left( \mathbf{ii}\right) $ like the
functional $\mathcal{S}_{4D}^{grav}\left[ \mathbf{E,\Omega }\right] $
reading as in eq(\ref{4sd}). The four dimensional $\mathbf{E}$ and $\mathbf{%
\Omega }$ are the vielbein and the spin connection 1-forms living on $%
\mathbb{R}^{2}\times \mathbb{CP}^{1}$ and valued in $sl(2,\mathbb{C})$.
These gravity potentials are given by linear combinations $(\mathbf{A}%
_{L}\pm \mathbf{A}_{R})/2$ \textrm{in a similar fashion to} the
Achucarro-Townsend 3D Chern- Simons derivation. We show that the two point
correlation functions $\left\langle \mathbf{E}(\zeta _{1})\mathbf{E}(\zeta
_{2})\right\rangle $ and $\left\langle \mathbf{\Omega }(\zeta _{1})\mathbf{%
\Omega }(\zeta _{2})\right\rangle $ vanish identically while $\left\langle
\mathbf{E}(\zeta _{1})\mathbf{\Omega }(\zeta _{2})\right\rangle $ is a non
vanishing propagator $\boldsymbol{P}(\zeta _{1}-\zeta _{2})$ given by eqs(%
\ref{pp}-\ref{pq}). \ \ \ \ \

Second, considering the above\ 4D topological field action (\ref{grav}) with
$\mathbf{E}$ and $\mathbf{\Omega }$ expressed in terms of CYW connections $(%
\mathbf{A}_{L}\pm \mathbf{A}_{R})/2$, we obtain two types of 4D
gravitational line defects W$_{E}\left[ \mathrm{\gamma }_{z}\right] $ and W$%
_{\Omega }\left[ \mathrm{\gamma }_{w}\right] $ described by topological
observables based on the holonomy of the topological gravitational 1-form
potentials
\begin{equation}
\Phi _{E}=\doint\nolimits_{\mathrm{\gamma }_{z}}\mathbf{E}\qquad ,\qquad
\Phi _{\Omega }=\doint\nolimits_{\mathrm{\gamma }_{w}}\mathbf{\Omega }
\label{fl}
\end{equation}%
with loops $\mathrm{\gamma }_{z}$ and $\mathrm{\gamma }_{w}$ spreading in
the topological plane $\mathbb{R}^{2}$ and located at the points z and w in $%
\mathbb{CP}^{1}\simeq \mathbb{S}^{2}$. Moreover, using the topological
invariant $\mathcal{S}_{4D}^{grav}$ as well as the vielbein and spin
connection line defects, we give partial results on quantum integrability
including Yang-Baxter and RLL equations; and comment\textrm{\ on the
extension of the AdS}$_{3}$\textrm{/CFT}$_{2}$\textrm{\ correspondence in
the new framework of 4D topological gravity (\ref{grav}).}

\textrm{Before proceeding, we would like to emphasise that the derivation of
the field action (\ref{grav}) of the 4D topological gravity will be carried
out in three consecutive steps \textbf{1}, \textbf{2} and \textbf{3} as
indicated by the following diagram in Table }\textbf{\ref{T}}.
\begin{table}[h]
\begin{center}
$%
\begin{tabular}{|ccccc|}
\hline\hline
&  &  &  &  \\
3D & : & AdS$_{3}$ gravity $S_{AdS_{3}}^{grav}\left[ e,\omega \right] $ & $\
\ \ \underrightarrow{\text{ \ \ step }1\text{ \ \ }}$ $\ \ \ $ & $S_{3D}^{cs}%
\left[ A_{L}\right] -S_{3D}^{cs}\left[ A_{R}\right] $ \\
&  &  &  &  \\
&  & $\downarrow $ $\ \ \text{steps }1$-$2$-$3$ &  & $\downarrow $ \ \ step $%
2$ \\
&  &  &  &  \\
4D & : & 4D gravity $S_{4D}^{grav}\left[ E,\Omega \right] $ & $%
\underleftarrow{\text{ \ \ step }3\text{ \ \ \ }}$ & $S_{4D}^{cyw}\left[
A_{L}\right] -S_{4D}^{cyw}\left[ A_{R}\right] $ \\
&  &  &  &  \\ \hline\hline
\end{tabular}%
$%
\end{center}
\par
\vspace{-0.5cm}
\caption{the way diagram to build 4D topological gravity by starting from AdS%
$_{3}$ gravity. The way is achieved into three steps \textbf{1}, \textbf{2}
and \textbf{3} as indicated by the arrows.}
\label{T}
\end{table}

The organisation of the paper is as follows: \textrm{In section 2}, we
revisit useful aspects \textrm{of} the Chern-Simons formulation of AdS$_{3}$
gravity and its boundary CFT$_{2}$. \textrm{In section 3}, we give the
building algorithm (BAL) for the derivation of the CWY theory by starting
from \textrm{the} usual 3D-Chern-Simons. \textrm{In section 4}, we develop
the basis of the 4D topological gravity deduced from the AdS$_{3}$ gravity
and investigate the topological defects \textrm{within}. \textrm{Section 5}
is devoted to conclusion and comments with regards to integrable quantum
spin chains and \textrm{the extension of} AdS$_{3}$/CFT$_{2}$ correspondence.

\section{ Chern-Simons formulation of AdS$_{3}$ gravity}

\qquad We begin by describing the Anti-de Sitter 3D gravity with spacetime
metric $g_{\mu \nu }=e_{\mu }^{a}\eta _{ab}e_{\nu }^{b}$ where $\eta _{ab}$
is the usual flat $diag(-,+,+)$ of 3D with rotation group in tangent space
given by $SO\left( 1,2\right) .$ This non compact space is a maximally
symmetric solution of 3D Einstein's equation with a negative cosmological
constant (negative constant curvature). By using the Dreibein $e_{\mu }^{a}$
and the (dualized) spin connection $\omega _{\mu }^{a}$ of AdS$_{3}$ as well
as the associated 1-forms $e^{a}=e_{\mu }^{a}dx^{\mu }$ and $\omega
^{a}=\omega _{\mu }^{a}dx^{\mu }$ invariant under Diff(AdS$_{3}$) but
transforming as $SO\left( 1,2\right) $ vectors, the field action $\mathcal{S}%
_{{\small 3D}}^{{\small grav}}=\mathcal{S}_{{\small 3D}}^{{\small grav}}%
\left[ e,\omega \right] $ of the AdS$_{3}$ gravity reads in differential
form language as follows \textrm{\cite{1E,2AA,2EA,3EA}}%
\begin{equation}
\mathcal{S}_{{\small 3D}}^{{\small grav}}=\frac{1}{8\pi G}\dint\nolimits_{%
\mathcal{M}_{3D}}\mathcal{L}_{3}^{grav}  \label{23}
\end{equation}%
The 3-form Lagrangian density reads in terms of the curvature 2-form $%
\mathcal{R}_{a}$ like,%
\begin{equation}
\begin{tabular}{lll}
$\mathcal{L}_{3}^{grav}$ & $=$ & $e^{a}\wedge \mathcal{R}_{a}+\frac{\xi }{3!}%
\varepsilon _{abc}e^{a}\wedge e^{b}\wedge e^{c}$ \\
$\mathcal{R}_{a}$ & $=$ & $d\omega _{a}+\frac{1}{2}\varepsilon _{abc}\omega
^{b}\wedge \omega ^{c}$%
\end{tabular}
\label{gr1}
\end{equation}%
where for convenience \textrm{we have set} $\xi =8\pi G_{N}/l_{{\small AdS}%
}^{2}$ with $l_{{\small AdS}}$ the AdS$_{3}$ radius and $G_{N}$ the Newton
constant in 3D. The field equations of motion following from the above $%
\mathcal{S}_{{\small 3D}}^{{\small grav}}$ are given by%
\begin{equation}
\begin{tabular}{lll}
$\mathcal{R}_{a}$ & $=$ & $-\frac{\xi }{2}\varepsilon _{abc}e^{b}\wedge
e^{c} $ \\
$de_{a}$ & $=$ & $-\frac{1}{2}\varepsilon _{abc}e^{b}\wedge \omega ^{c}$%
\end{tabular}%
\end{equation}%
and are solved by the AdS$_{3}$ metric \textrm{\cite{ADSM}}. Notice that the
3D field action $\mathcal{S}_{{\small 3D}}^{{\small grav}}$ is sometimes
termed as the spin $s_{cft}=2$ AdS$_{3}$ gravity; a terminology due to the
AdS$_{3}$/CFT$_{2}$ correspondence \textrm{\cite{2AA,Vero,BEN}}. This
duality means that on the boundary of AdS$_{3}$ lives a scale invariant
field theory with quantum states described by conformal symmetry generated
by a conformal spin $s_{cft}=2$ current T$_{zz}$ satisfying the Virasoro
algebra \textrm{\cite{5A}.}

On the other hand, following \textrm{\cite{1E,2AA,2EAB,5A}}, the gravity
field action $\mathcal{S}_{{\small 3D}}^{{\small grav}}$ given by (\ref{gr1}%
) can be expressed like the difference of two Chern-Simons field actions as
follows%
\begin{equation}
\mathcal{S}_{{\small 3D}}^{{\small grav}}=\mathcal{S}_{{\small 3D}}^{{\small %
CS}}\left[ A_{L}\right] -\mathcal{S}_{{\small 3D}}^{{\small CS}}\left[ A_{R}%
\right]  \label{1}
\end{equation}%
with
\begin{equation}
\mathcal{S}_{{\small 3D}}^{{\small CS}}\left[ A\right] =\frac{k}{4\pi }%
\dint\nolimits_{\mathcal{M}_{3D}}Tr\left( AdA+\frac{2}{3}A^{3}\right)
\label{cs}
\end{equation}%
In the eqs(\ref{cs}), the Chern-Simons levels $k_{L}$ and $k_{R}$ are taken
equal; and the $A_{L}=A_{L}(x^{0},x^{1},x^{2})$ and $%
A_{R}=A_{R}(x^{0},x^{1},x^{2})$ are the 3D Chern-Simons gauge potentials.
They are valued in the $SO\left( 1,2\right) _{L}$ and $SO\left( 1,2\right)
_{R}$ gauge symmetries having the homomorphisms%
\begin{equation}
SO\left( 1,2\right) \simeq SU\left( 1,1\right) \simeq SL\left( 2,\mathbb{R}%
\right)  \label{sss}
\end{equation}%
These Chern-Simons 1-forms $A_{L}$ and $A_{R}$ expand in terms of the
space-time 1-forms $dx^{\mu }$ and the three $SU\left( 1,1\right) $
generators $J_{a}$ like
\begin{equation}
A=A^{a}J_{a}=dx^{\mu }A_{\mu }=dx^{\mu }A_{\mu }^{a}J_{a}  \label{AA}
\end{equation}

The link between the CS gauge fields $A_{L},$ $A_{R}$ and the gravity fields
$e_{\mu }^{a},$ $\omega _{\mu }^{a}$ is given by the Achucarro-Townsend
relations \textrm{\cite{1E,2AA,2EAB}}%
\begin{equation}
\begin{tabular}{lll}
$A_{L}^{a}$ & $=$ & $\omega ^{a}+\frac{1}{l_{{\small AdS}}}e^{a}$ \\
$A_{R}^{a}$ & $=$ & $\omega ^{a}-\frac{1}{l_{{\small AdS}}}e^{a}$%
\end{tabular}
\label{14}
\end{equation}%
In what follows, we set $l_{{\small AdS}}=1$ for simplicity; and by using eq(%
\ref{AA}), we then have $A_{L/R}^{a}=\omega ^{a}\pm e^{a}$. With the
relations (\ref{23}-\ref{gr1}) and (\ref{1}-\ref{cs}) as well as (\ref{14}),
we have completed the step-\textbf{1} in the way diagram of Table \textbf{%
\ref{T}}.

\section{From 3D Chern-Simons to CWY theory}

\qquad \textrm{First, we recall useful aspects of the 3D Chern-Simons
theory; then we give the Building ALgorithm (BAL) where we briefly describe
the main pillars to extend the 3D Chern-Simons theory to the CWY theory.
This investigation constitutes the second step of Table \textbf{\ref{T}}
towards the derivation of the novel topological 4D gravity (\ref{grav}).}

\subsection{More on 3D Chern-Simons action}

\qquad In 3D, the Chern-Simons gauge field action $\mathcal{S}_{3D}^{{\small %
CS}}=\int_{\mathcal{M}_{3D}}\mathcal{L}_{3}^{{\small CS}}$ with generic
gauge symmetry G\ depends on the 1-form gauge potential field $dx^{\mu
}A_{\mu }^{a}$ as follows \textrm{\cite{DSA}}%
\begin{equation}
\mathcal{L}_{3}^{{\small CS}}=A^{a}\wedge dA_{a}+\frac{2}{3}\varepsilon
_{abc}A^{a}\wedge A^{b}\wedge A^{c}  \label{3d}
\end{equation}%
Here, we will think about G either as $SU\left( 2\right) $ or like $SU\left(
1,1\right) $ having the homomorphisms (\ref{sss}). These two gauge
symmetries are the two real forms of the complex $SL(2,\mathbb{C})$ which
turns out to play a basic role in our derivation of 4D topological gravity.
\textrm{In passing, note that when considering }applications in 3D higher
spin theory, the tangent space of the 3D spacetime can be imagined either as
the euclidian $\mathbb{R}^{3}$ with $SO\left( 3\right) \simeq SU\left(
2\right) $ isotropy, or like the Lorentzian $\mathbb{R}^{1,2}$ with $%
SO\left( 1,2\right) \simeq SU\left( 1,1\right) $ symmetry.

The field equation of the three dimensional $SL\left( 2,\mathbb{R}\right) $
CS gauge field is given by%
\begin{equation}
F_{a}=dA_{a}+\varepsilon _{abc}A^{b}\wedge A^{c}=0
\end{equation}%
By using the expansions $A\left( x\right) =A^{a}\left( x\right) J_{a}$\ and $%
F\left( x\right) =F^{a}\left( x\right) J_{a}$ in terms of the generators $%
J_{a}$ of the Lie algebra of the gauge symmetry, the above field equation
reads shortly as follows%
\begin{equation}
F=dA+A\wedge A=0  \label{2f}
\end{equation}%
So, the CS gauge field $A$ has a flat curvature showing that the 3D
Chern-Simons theory has no observable constructed out of the 2-form $F.$
However, one can still build gauge invariant observables in this topological
3D gauge theory; they are given by gauge topological line defects such as
the Wilson loops $\mathrm{\gamma }$ on which propagate quantum states $%
\left\vert \lambda \right\rangle $ sitting in some representation $%
\boldsymbol{R}$ of the gauge symmetry G. These observables are built as
\textrm{\cite{1A},}%
\begin{equation}
W_{\boldsymbol{R}}\left[ \mathrm{\gamma }\right] =Tr_{\boldsymbol{R}}\left[
P\exp \left( \doint\nolimits_{\mathrm{\gamma }}A\right) \right]
\end{equation}%
where for the case $G=SU\left( 2\right) $, the $\boldsymbol{R}$
representation (labeled as $\boldsymbol{R}_{j}$) have spin states $%
\left\vert j,m\right\rangle $ sitting in multiplets $\mathfrak{M}_{j}$ with
dimension $2j+1.$ This construction extends straightforwardly to the
classical gauge symmetries in the Cartan classification of Lie algebras; but
here below we will focuss on the real forms of $SL\left( 2,\mathbb{C}\right)
$.

\subsection{CWY theory: building algorithm}

\qquad The \textrm{CWY theory} is a 4D topological quantum field theory that
can derived from the usual 3D Chern-Simons theory using BAL\textrm{\cite{1A}%
. This topological QFT}$_{4D}$\textrm{\ also has a six dimensional origin
\cite{6DO}}. From a practical point of view, the CWY construction is a
complexified gauge theory living on the 4D space $\boldsymbol{M}_{4}=\Sigma
_{2}\times C$ with $\Sigma _{2}$ a real surface thought of here as $\mathbb{R%
}^{2}$; and $C$ a complex curve which can be either $\left( i\right) $ the
complex projective line $\mathbb{CP}^{1}$, $\left( ii\right) $ the complex
line $\mathbb{C}^{\times }$ without the origin; or $\left( iii\right) $ an
elliptic curve $\mathcal{E}$. This theory has been a subject of big interest
in the few last years especially when it comes to quantum integrability and
brane realisation of integrable systems \textrm{\cite{1D,1DA,1DB,2EB}}.
\newline

Below, we give a rough re-derivation of this theory\textrm{; }starting from
eq(\ref{3d}) and following \textrm{\cite{1A}}, one can construct a 4D
extension of the usual 3D Chern-Simons gauge theory by performing some
surgery on the 3D Chern-Simons theory. The building algorithm of the 4D
topological CWY theory from 3D-Chern-Simons may be done into three steps as
follows.

\begin{description}
\item[$\left( \mathbf{1}\right) $] the 3D space $\mathcal{M}_{3D}$ in the CS
field action
\begin{equation}
\mathcal{S}_{3D}^{{\small CS}}=\int_{\mathcal{M}_{3D}}\mathcal{L}_{3}^{%
{\small CS}}  \label{b1}
\end{equation}%
is broken to $R_{t}\times \Sigma _{2}$. The surface $\Sigma _{2}$ is named
the real \emph{topological plane} where \textrm{spread} the gauge invariant
line defects $\mathrm{\gamma }$ of the CWY theory. The real line $\mathrm{%
\gamma }$ can be imagined in terms of sites in the integrable quantum spin
chain; or as branes intersection line in type II strings and M- theory
\textrm{\cite{1D}-\cite{1DB}}. In what follows, this surface $\Sigma _{2}$
will be taken as $\mathbb{R}^{2}$.

\item[$\mathbf{(2)}$] promote the time axis $R_{t}$ to a complex line $C$
with local coordinate $z=x^{3}+it$; this is possible by adding a fourth
coordinate $x^{3}$ to the old three ($t,x^{1},x^{2}$); thus leading to ($%
t,x^{1},x^{2},x^{3}$) imagined as ($x,y,z,\bar{z}$). Here, we think about
this $z$ as the local coordinate of the projective $\mathbb{CP}^{1}$ which
is isomorphic to the real 2-sphere $\mathbb{S}^{2}.$ The complex line $C$ in
the CWY theory is sometimes termed as the \emph{holomorphic plane}. With
this surgery, the space $\mathcal{M}_{3D}$ of the Chern-Simons theory
becomes a 4D space $\boldsymbol{M}_{4}$ factorised like
\begin{equation}
\boldsymbol{M}_{4}=\Sigma _{2}\times \mathbb{CP}^{1}=\mathbb{R}^{2}\times
\mathbb{S}^{2}
\end{equation}%
Regarding the fields and symmetries of the CWY theory, we have the following
emerging quantities: $\left( i\right) $ Under surgery, the initial 1-form 3D
CS gauge field $A^{a}\left( x,y,t\right) $ with expansion $%
A^{a}=A_{x}^{a}dx+A_{y}^{a}dy+A_{t}^{a}dt$ becomes a complex 4D gauge field $%
\boldsymbol{A}^{a}=\boldsymbol{A}^{a}\left( x,y,z,\bar{z}\right) $ with the
expansion%
\begin{equation}
dz\wedge \boldsymbol{A}^{a}=d\bar{z}\wedge \left( \boldsymbol{A}_{x}^{a}dx+%
\boldsymbol{A}_{y}^{a}dy+\boldsymbol{A}_{\bar{z}}^{a}d\bar{z}\right)
\label{cy}
\end{equation}%
that we denote shortly as\textrm{\ }$dz\wedge d\zeta ^{\text{\textsc{a}}}A_{%
\text{\textsc{a}}}$ where $\zeta ^{\text{\textsc{a}}}$\ refers to ($x,y,\bar{%
z}$)\textrm{.} $\left( ii\right) $ the gauge symmetry, which in the 3D
Chern-Simons was taken as $su\left( 2\right) $ or $su(1,1)$, gets replaced
by the complexified version namely $sl(2,\mathbb{C})$.

\item[$\left( \mathbf{3}\right) $] the gauge field action $\mathcal{S}_{4D}^{%
{\small CWY}}$ resulting from the promotion of the 3D CS theory to $%
\boldsymbol{M}_{4}=\mathbb{R}^{2}\times \mathbb{CP}^{1}$ defines the CWY
theory. This field action is complex and can be presented as follows
\begin{equation}
\mathcal{S}_{4D}^{{\small CWY}}=\dint\nolimits_{\boldsymbol{M}_{4}}\mathcal{L%
}_{4}^{{\small CWY}}
\end{equation}%
with Lagrangian 4-form given by the trace%
\begin{equation}
\mathcal{L}_{4}^{{\small CWY}}=dz\wedge Tr\left( \boldsymbol{A}^{a}\wedge d%
\boldsymbol{A}_{a}+\frac{2}{3}\varepsilon _{abc}\boldsymbol{A}^{a}\wedge
\boldsymbol{A}^{b}\wedge \boldsymbol{A}^{c}\right)   \label{LCWY}
\end{equation}%
Notice that because of the factor $dz\wedge $ in the Lagrangian 4-form, the
1-form gauge field $\boldsymbol{A}^{a}\left( x,y,z,\bar{z}\right) $ living
on $\boldsymbol{M}_{4}$ contributes only through the partial expansion (\ref%
{cy}) namely $\boldsymbol{A}_{x}^{a}dx+\boldsymbol{A}_{y}^{a}dy+A_{\bar{z}%
}^{a}d\bar{z}$. The shift of $\boldsymbol{A}^{a}$ by the missing term $%
\boldsymbol{A}_{z}^{a}dz$ is a symmetry of $\mathcal{L}_{4}$ and so can be
dropped out due to the property $dz\wedge dz=0$. \newline
\end{description}

\begin{figure}[tbph]
\begin{center}
\includegraphics[width=16cm]{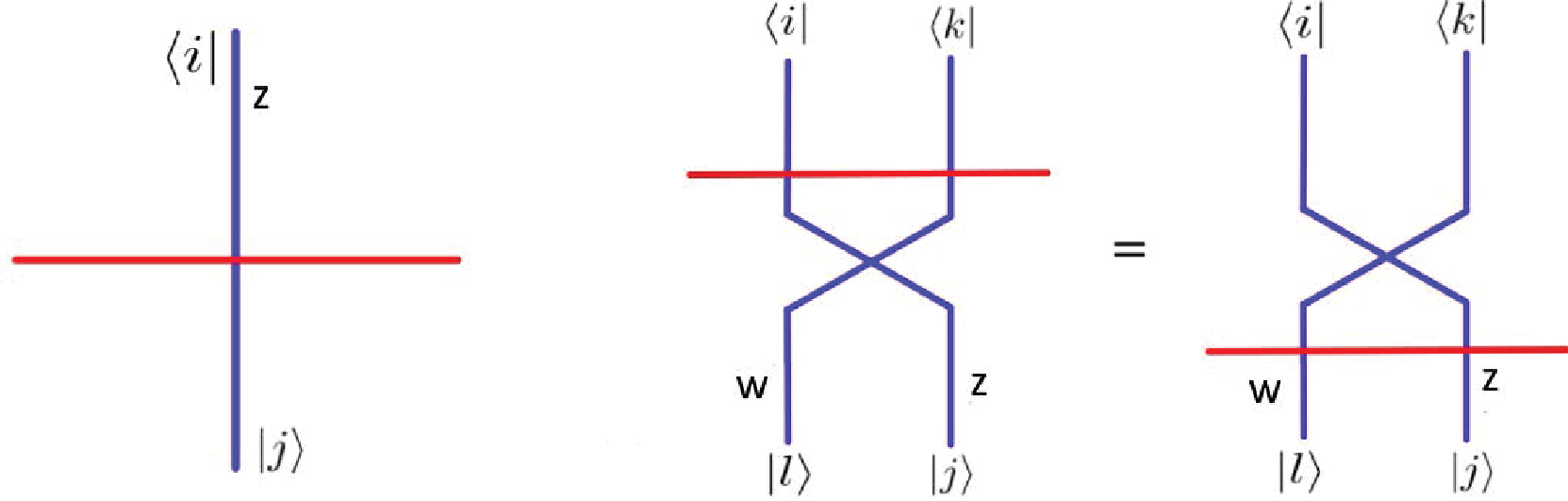}
\end{center}
\par
\vspace{-0.5cm}
\caption{On the left, the Lax operator $\mathcal{L}\left( z\right) $ given
by the crossing of a 't Hooft line at z=0 (in red) and a Wilson line at z
(in blue) with incoming $\left\langle i\right\vert $ and out going $%
\left\vert j\right\rangle $ states. On the right, the RLL relations encoding
the commutation relations between two L-operators at z and w.}
\label{M1}
\end{figure}

The field equation of motion of the CWY gauge potential (\ref{cy}) expressed
as $\boldsymbol{A}=\boldsymbol{A}^{a}J_{a}$ is given by%
\begin{equation}
H_{3}=dz\wedge \boldsymbol{F}_{2}=0\qquad \Rightarrow \qquad \boldsymbol{F}%
_{2}=0  \label{f2}
\end{equation}%
with $\boldsymbol{F}_{2}=d\boldsymbol{A}+\boldsymbol{A}\wedge \boldsymbol{A}$
expanding like $d\zeta ^{\text{\textsc{a}}}\wedge d\zeta ^{\text{\textsc{b}}%
}F_{\left[ \text{\textsc{ab}}\right] }.$ So, for this 4D Chern-Simons theory
there is no observable constructed from the gauge curvature $\boldsymbol{F}%
_{2}$. However, we do have topological observables given by surface and line
defects \textrm{\cite{1A,2A,1B,1BC} }as exemplified by the pictures of the
Figure \textbf{\ref{M1}}. For instance, we have the Wilson loop $W_{%
\boldsymbol{R}}\left[ \mathrm{\gamma }_{z}\right] $ defined as%
\begin{equation}
W_{\boldsymbol{R}}\left[ \mathrm{\gamma }_{z}\right] =P\exp \left(
\doint\nolimits_{\mathrm{\gamma }}\boldsymbol{A}\right) =P\exp \left(
\doint\nolimits_{\mathrm{\gamma }}\left( \boldsymbol{A}_{x}dx+\boldsymbol{A}%
_{y}dy\right) \right)  \label{W}
\end{equation}%
with loop $\mathrm{\gamma }_{z}$ spreading in the topological plane $\mathbb{%
R}^{2}$; but located the point z in $\mathbb{CP}^{1}$. The prefactor $P$
refers to path ordering.
Notice that in eq(\ref{W}), the usual trace has been dropped out while $W_{%
\boldsymbol{R}}\left[ \mathrm{\gamma }_{z}\right] $ still preserving gauge
symmetry; this remarkable feature is due to asymptotic conditions described
in \textrm{\cite{1A}}. Along with the line defects, one may also have their
couplings given by lines' crossings like those involved in Yang-Baxter
equation (YBE). Another special type of lines crossings is given by the
so-called Lax operator denoted as $\boldsymbol{L}_{(\mathbf{\hat{\mu})}}$;
it describes the crossing of Wilson line $W_{\boldsymbol{R}}\left[ \mathrm{%
\gamma }_{z}\right] $ by a 't Hooft line tH$_{\boldsymbol{\mu }}\left[
\mathrm{\gamma }_{0}\right] $ characterised by a minuscule coweight $\mathbf{%
\hat{\mu}}.$ An interesting realisation of this $\boldsymbol{L}_{(\mathbf{%
\hat{\mu})}}$ has been obtained in the CWY theory; it is given by \textrm{%
\cite{1B},}%
\begin{equation}
\boldsymbol{L}_{\mathbf{\mu }}=e^{X}z^{\mathbf{\hat{\mu}}}e^{Y}  \label{lm}
\end{equation}%
where $\mathbf{\hat{\mu}}$ is a minuscule coweight and X and Y are nilpotent
operators of the gauge symmetry of the CWY theory. For explicit calculations
regarding eq(\ref{lm}), see \textrm{\cite{1DB}}. With this description
regarding lines defects and their couplings, we complete the step-\textbf{2}
in the way diagram of Table \textbf{\ref{T}}.

\section{From AdS$_{3}$ to topological 4D gravity}

\qquad In this section, we\textrm{\ carry out the third step} in the way
diagram of Table \textbf{\ref{T}}. First, we derive the field action (\ref%
{grav}) of the topological 4D gravity from the AdS$_{3}$ theory by \textrm{%
implementing} the building algorithm used in the derivation of the CWY
theory as described above. Then, we construct gravitational line defects
\textrm{of} the 4D topological theory and \textrm{discuss their} application
in integrable quantum system\textrm{s}.

\subsection{Deriving the topological 4D gravity}

\qquad Following \textrm{\cite{1E,2AA}}, the AdS$_{3}$ gravity can be
formulated as a difference of two Chern-Simons gauge theor\textrm{ies}. In
this tricky formulation briefly revisited in section 2 as shown by eqs(\ref%
{1}-\ref{cs}), the AdS$_{3}$ gravity action reads in terms of\textrm{\ the}
differential- form language as the difference of two 3D CS- field actions as
follows%
\begin{equation}
\mathcal{S}_{3D}^{gravity}=\dint\nolimits_{\mathcal{M}_{3D}}\mathcal{L}_{3}^{%
{\small CS}_{L}}-\dint\nolimits_{\mathcal{M}_{3D}}\mathcal{L}_{3}^{{\small CS%
}_{R}}  \label{3gr}
\end{equation}%
where $\mathcal{L}_{3}^{{\small CS}_{L}}$ and $\mathcal{L}_{3}^{{\small CS}%
_{R}}$ are Chern-Simons 3-forms with gauge symmetries G$_{L}$ and G$_{R}$.

\subsubsection{Action $\mathcal{S}_{4D}^{gravity}$ in terms of CWY fields}

\qquad Applying the building algorithm, used in the derivation of the CWY
theory (\ref{b1}-\ref{f2}), to the 3D AdS$_{3}$ gravity action (\ref{3gr}),
we end up with a 4D topological gravity described by%
\begin{equation}
\mathcal{S}_{4D}^{gravity}=\dint\nolimits_{\boldsymbol{M}_{4D}}\mathcal{L}%
_{4}^{gravity}  \label{4dg}
\end{equation}%
with%
\begin{equation}
\mathcal{L}_{4}^{gravity}=\mathcal{L}_{4}^{{\small CWY}_{L}}-\mathcal{L}%
_{4}^{{\small CWY}_{R}}
\end{equation}%
In this field action, the 4D space $\boldsymbol{M}_{4D}$ is given by the
fibration $\mathbb{R}^{2}\times \mathbb{CP}^{1}$ while the left $\mathcal{L}%
_{4}^{{\small CWY}_{L}}$ and the right $\mathcal{L}_{4}^{{\small CWY}_{R}}$
Lagrangian 4-forms read as follows%
\begin{equation}
\begin{tabular}{lll}
$\mathcal{L}_{4}^{{\small CWY}_{L}}$ & $=$ & $dz\wedge tr\mathcal{L}_{3}^{%
{\small CWY}_{L}}$ \\
$\mathcal{L}_{4}^{{\small CWY}_{R}}$ & $=$ & $dz\wedge tr\mathcal{L}_{3}^{%
{\small CWY}_{R}}$%
\end{tabular}%
\end{equation}%
with
\begin{eqnarray}
\mathcal{L}_{3}^{{\small CWY}_{L}} &=&\boldsymbol{A}_{L}\wedge d\boldsymbol{A%
}_{L}+\frac{2}{3}\boldsymbol{A}_{L}\wedge \boldsymbol{A}_{L}\wedge
\boldsymbol{A}_{L}  \label{l} \\
\mathcal{L}_{3}^{{\small CWY}_{R}} &=&\boldsymbol{A}_{R}\wedge d\boldsymbol{A%
}_{R}+\frac{2}{3}\boldsymbol{A}_{R}\wedge \boldsymbol{A}_{R}\wedge
\boldsymbol{A}_{R}  \label{r}
\end{eqnarray}%
Notice that the gauge fields $\boldsymbol{A}_{L/R}=\boldsymbol{A}%
_{L/R}\left( x,y,z,\bar{z}\right) $ are given by eq(\ref{cy}); i.e $%
\boldsymbol{A}_{L/R}=d\zeta ^{\text{\textsc{a}}}A_{\text{\textsc{a}}}^{L/R}$%
; and the operator $dz\wedge $ is as in the CWY topological theory. Here,
the $\boldsymbol{A}_{L}$ is valued into the Lie algebra $sl(2,\mathbb{C})_{L}
$ generated by $J_{L}^{a}$; that is $\boldsymbol{A}_{L}=J_{a}^{L}A_{L}^{a}$.
Similarly, the $\boldsymbol{A}_{R}$ is valued into the Lie algebra $sl(2,%
\mathbb{C})_{R}$ generated by $J_{R}^{a}$ allowing $\boldsymbol{A}%
_{R}=J_{a}^{R}A_{R}^{a}.$ By performing the trace in the above
relationships, we obtain%
\begin{eqnarray}
\mathcal{L}_{3}^{{\small CWY}_{L}} &=&q_{ab}\boldsymbol{A}_{L}^{a}\wedge d%
\boldsymbol{A}_{L}^{b}+\frac{2}{3}\varepsilon _{abc}\boldsymbol{A}%
_{L}^{a}\wedge \boldsymbol{A}_{L}^{b}\wedge \boldsymbol{A}_{L}^{c}  \label{L}
\\
\mathcal{L}_{3}^{{\small CWY}_{R}} &=&q_{ab}\boldsymbol{A}_{R}^{a}\wedge d%
\boldsymbol{A}_{R}^{b}+\frac{2}{3}\varepsilon _{abc}\boldsymbol{A}%
_{R}^{a}\wedge \boldsymbol{A}_{R}^{b}\wedge \boldsymbol{A}_{R}^{c}  \label{R}
\end{eqnarray}%
with Killing form $q_{ab}=Tr\left( J_{a}J_{b}\right) $ for both $sl(2,%
\mathbb{C})_{L}$ and $sl(2,\mathbb{C})_{R}.$ The field equation of the CWY
gauge potentials 1-forms $\boldsymbol{A}_{L}$ and $\boldsymbol{A}_{R}$ are
given by%
\begin{eqnarray}
\boldsymbol{H}_{3}^{L} &=&dz\wedge \boldsymbol{F}_{2}^{L}=0\,\qquad
\Rightarrow \qquad \boldsymbol{F}_{2}^{L}=0  \label{f2l} \\
\boldsymbol{H}_{3}^{R} &=&dz\wedge \boldsymbol{F}_{2}^{R}=0\,\qquad
\Rightarrow \qquad \boldsymbol{F}_{2}^{R}=0  \label{f2r}
\end{eqnarray}%
where%
\begin{eqnarray}
\boldsymbol{F}_{2}^{L} &=&d\boldsymbol{A}_{L}+\boldsymbol{A}_{L}\wedge
\boldsymbol{A}_{L} \\
\boldsymbol{F}_{2}^{R} &=&d\boldsymbol{A}_{R}+\boldsymbol{A}_{R}\wedge
\boldsymbol{A}_{R}
\end{eqnarray}%
The topological 4D gravity action (\ref{4dg}) is one of the main results in
this paper. Its explicit expression in terms\textrm{\ of the} 4D gravity
1-forms is obtained by substituting into eq(\ref{4dg}) the $A_{L/R}^{a}$ by
the following
\begin{equation}
\begin{tabular}{lll}
$A_{L}^{a}$ & $=$ & $\Omega ^{a}+E^{a}$ \\
$A_{R}^{a}$ & $=$ & $\Omega ^{a}-E^{a}$%
\end{tabular}
\label{EW}
\end{equation}%
They behave as 3-vectors of the complexified gauge symmetry; that is $%
A_{L}^{a}$ a complex triplet of $SL(2,\mathbb{C})_{L}$ while $A_{R}^{a}$ a
triplet of $SL(2,\mathbb{C})_{R}$. \textrm{In this new setting}, the complex
potential $E^{a}$ is the 1-form dreibein expanding as $d\zeta ^{\text{%
\textsc{a}}}E_{\text{\textsc{a}}}^{a}$ with sections $E_{\text{\textsc{a}}%
}^{a}=E_{\text{\textsc{a}}}^{a}\left( x,y,z,\bar{z}\right) $ living in the
4D space $\mathbb{R}^{2}\times \mathbb{CP}^{1}$; and the complex $\Omega ^{a}
$ is the l-form spin connection $d\zeta ^{\text{\textsc{a}}}\Omega _{\text{%
\textsc{a}}}^{a}$ with components $\Omega _{\text{\textsc{a}}}^{a}=\Omega _{%
\text{\textsc{a}}}^{a}\left( x,y,z,\bar{z}\right) $. For later use, we
denote the two- points Green functions of these gauge fields $%
A_{L/R}^{a}\left( \zeta \right) =d\zeta ^{\text{\textsc{a}}}A_{\text{\textsc{%
a}}L/R}^{a}\left( \zeta \right) $ as follows%
\begin{equation}
\begin{tabular}{lll}
$\left\langle A_{\text{\textsc{a}}L}^{a}\left( \zeta _{1}\right) A_{\text{%
\textsc{b}}L}^{b}\left( \zeta _{2}\right) \right\rangle $ & $=$ & $\mathcal{P%
}_{\text{\textsc{ab}}L}^{ab}\left( \zeta _{1}-\zeta _{2}\right) $ \\
$\left\langle A_{\text{\textsc{a}}L}^{a}\left( \zeta _{1}\right) A_{\text{%
\textsc{b}}R}^{b}\left( \zeta _{2}\right) \right\rangle $ & $=$ & $0$ \\
$\left\langle A_{\text{\textsc{a}}R}^{a}\left( \zeta _{1}\right) A_{\text{%
\textsc{b}}R}^{b}\left( \zeta _{2}\right) \right\rangle $ & $=$ & $\mathcal{P%
}_{\text{\textsc{ab}}R}^{ab}\left( \zeta _{1}-\zeta _{2}\right) $%
\end{tabular}
\label{3A}
\end{equation}%
where the translation invariant tensor $\mathcal{P}_{\text{\textsc{ab}}}^{ab}
$ can be read from \textrm{\cite{1A,2A}}. Because of the minus sign in the $%
\mathcal{L}_{4}^{gravity}$ gravity Lagrangian given by the difference $%
\mathcal{L}_{4}^{{\small CWY}_{L}}-\mathcal{L}_{4}^{{\small CWY}_{R}}$, the
two-points Green functions in the right sector are related to their
homologue in the left sector as
\begin{equation}
\mathcal{P}_{\text{\textsc{ab}}R}^{ab}=-\mathcal{P}_{\text{\textsc{ab}}%
L}^{ab}\equiv -\mathcal{P}_{\text{\textsc{ab}}}^{ab}  \label{3p}
\end{equation}

\subsubsection{ Action $\mathcal{S}_{4D}^{gravity}$ in terms of gravity
fields}

\qquad Here, we study the building of the action $\mathcal{S}_{4D}^{gravity}$
in terms of 4D gravity fields $E^{a}$ and $\Omega ^{a};$ it can be obtained
by substituting (\ref{EW}) into eqs (\ref{4dg}-\ref{R}). From the relations (%
\ref{EW}), we deduce interesting features; in particular the \textrm{%
following two}:

$\left( \mathbf{1}\right) $ The complexified 1-form spin connection $\Omega
^{a}$ is related to the left and right CWY fields as the mean field of the
two gauge potentials namely $(A_{L}^{a}+A_{R}^{a})/2$; while the
complexified 1-form vielbein $E^{a}$ is given by the reduced field given by $%
(A_{L}^{a}-A_{R}^{a})/2$. Notice that \textrm{by} putting $%
A_{L}^{a}=A_{R}^{a}=A^{a}$; then the $\left. E^{a}\right\vert
_{A_{L}^{a}=A_{R}^{a}}$ vanishes and the spin connection $\left. \Omega
^{a}\right\vert _{A_{L}^{a}=A_{R}^{a}}$ reduces to $A^{a}$; thus leading to
\begin{equation}
\left. \mathcal{L}_{4}^{{\small CWY}_{L}}\right\vert
_{A_{L}^{a}=A_{R}^{a}}=\left. \mathcal{L}_{4}^{{\small CWY}_{R}}\right\vert
_{A_{L}^{a}=A_{R}^{a}}
\end{equation}
and consequently $\left. \mathcal{S}_{4D}^{gravity}\right\vert
_{A_{L}^{a}=A_{R}^{a}}$ vanishes. This property indicates that the
topological Lagrangian 4-form $\mathcal{L}_{4}^{gravity}(\mathbf{E},\mathbf{%
\Omega })$ is factorised like
\begin{equation}
\mathcal{S}_{4D}^{gravity}=\dint\nolimits_{\boldsymbol{M}_{4D}}\mathcal{L}%
_{4}^{gravity}=\dint\nolimits_{\boldsymbol{M}_{4D}}E_{a}\wedge \mathcal{K}%
_{3}^{a}  \label{ek}
\end{equation}%
with $\mathcal{K}_{3}=\mathcal{K}_{3}(\mathbf{E},\mathbf{\Omega })$ is a
3-form function of the vielbein $\mathbf{E}$ and the spin connection $%
\mathbf{\Omega }$ one-forms.

$\left( \mathbf{2}\right) $ Using the operators $J^{a}=(J_{L}^{a}+J_{R}^{a})/%
\sqrt{2}$ generating the diagonal Lie algebra of the CS gauge symmetry namely%
\begin{equation}
SL(2,\mathbb{C})_{+}=\frac{SL(2,\mathbb{C})_{L}\times SL(2,\mathbb{C})_{R}}{%
SL(2,\mathbb{C})_{-}}  \label{dia}
\end{equation}%
one can deal with the vielbein $\mathbf{E}$ and the spin connection $\mathbf{%
\Omega }$ as 1-form matrices valued in $sl(2,\mathbb{C})_{+};$ thus
facilitating the explicit calculations. In this formulation, the two 1-forms
decompose like $\mathbf{\Omega }=\Omega ^{a}J_{a}$ and $\mathbf{E}%
=E^{a}J_{a} $; which by using the Killing form $q_{ab}=tr\left(
J_{a}J_{b}\right) $, give $tr\left( J_{a}\mathbf{\Omega }\right)
=q_{ab}\Omega ^{b}$ and $tr\left( J_{a}\mathbf{E}\right) =q_{ab}E^{b}.$ With
the help of the inverse matrix $q^{ba}$, we also have
\begin{equation}
\Omega ^{a}=q^{ab}tr\left( J_{b}\mathbf{\Omega }\right) \qquad ,\qquad
E^{a}=q^{ab}tr\left( J_{b}\mathbf{E}\right)
\end{equation}%
consequently the eq(\ref{ek}) can be rewritten as%
\begin{equation}
\mathcal{S}_{4D}^{gravity}=\dint\nolimits_{\boldsymbol{M}_{4D}}tr\left(
\mathbf{E}\wedge \mathbf{K}_{3}\right)  \label{EK3}
\end{equation}%
Moreover, because of the building algorithm (\ref{b1}-\ref{f2}), the gravity
1-forms have partial expansions as follows%
\begin{eqnarray}
E^{a} &=&E_{x}^{a}dx+E_{y}^{a}dy+E_{\bar{z}}^{a}d\bar{z}  \label{ee} \\
\Omega ^{a} &=&\Omega _{x}^{a}dx+\Omega _{y}^{a}dy+\Omega _{\bar{z}}^{a}d%
\bar{z}  \label{ww}
\end{eqnarray}%
with no component $E_{z}^{a}dz$ nor $\Omega _{z}^{a}dz$. Combining these
quantities with eq(\ref{gr1}), we end up with the expression of the
topological 4D gravity in terms of the complexified vielbein and the spin
connection namely
\begin{equation}
\begin{tabular}{lll}
$\mathcal{L}_{4}^{gravity}$ & $=$ & $dz\wedge \left( E^{a}\wedge \boldsymbol{%
R}_{a}+\xi \varepsilon _{abc}E^{a}\wedge E^{b}\wedge E^{c}\right) $ \\
$\boldsymbol{R}_{a}$ & $=$ & $d\Omega _{a}+\frac{1}{2}\varepsilon
_{abc}\Omega ^{b}\wedge \Omega ^{c}$%
\end{tabular}
\label{419}
\end{equation}%
\textrm{In comparison} with (\ref{ek},\ref{EK3}), the 3-form $\mathbf{K}_{3}$
is given by%
\begin{equation}
\mathbf{K}_{3}=dz\wedge \left( \boldsymbol{R}+\xi \mathbf{E}\wedge \mathbf{E}%
\right)
\end{equation}%
Furthermore, using the formulation (\ref{4dg}) of the obtained 4D
topological gravity, one can \textrm{compute} interesting quantities
characterising \textrm{the} $\mathcal{S}_{4D}^{grav};$ f\textrm{or instance,
we can construct observables using} the gauge $\mathbf{A}_{L}$ and $\mathbf{A%
}_{R}$ (\textrm{or} equivalently the gravitational $\mathbf{\Omega }$ and $%
\mathbf{E}$) through their gauge invariant holonomies \textrm{as shown} in
next subsection. In due time, notice that\textrm{\ }by using the short
notation $A=d\zeta ^{\text{\textsc{a}}}A_{\text{\textsc{a}}}$\ introduced in
eq(\ref{cy}) while setting $U=d\zeta ^{\text{\textsc{a}}}U_{\text{\textsc{a}}%
}$\ and\textrm{\ }$\boldsymbol{V}=d\zeta ^{\text{\textsc{a}}}V_{\text{%
\textsc{a}}}$ with $\boldsymbol{U}$ and $\boldsymbol{V}$ \textrm{referring to%
} $\boldsymbol{\Omega }$ and $\boldsymbol{E}$; then calculating their
product, we obtain
\begin{equation}
\boldsymbol{UV}=\frac{1}{2}d\zeta ^{\text{\textsc{b}}}\wedge d\zeta ^{\text{%
\textsc{c}}}Z_{\text{\textsc{bc}}}\qquad ,\qquad Z_{\text{\textsc{bc}}}=%
\left[ U_{\text{\textsc{b}}},V_{\text{\textsc{c}}}\right]  \label{uv}
\end{equation}%
For the case where both $\boldsymbol{U}$ and $\boldsymbol{V}$ \textrm{are
equal to} $\mathbf{A}_{L}$ or to $\mathbf{A}_{R}$, the commutators $\left[
A_{L\text{\textsc{b}}},A_{L\text{\textsc{c}}}\right] $ and $\left[ A_{R\text{%
\textsc{b}}},A_{R\text{\textsc{c}}}\right] $ are respectively given by $%
A_{L}^{b}\mathrm{f}_{bc}^{a}A_{L\text{\textsc{c}}}^{c}$ and $A_{R}^{b}%
\mathrm{\bar{f}}_{bc}^{a}A_{R\text{\textsc{c}}}^{c}$\ with $\mathrm{f}%
_{bc}^{a}$ and $\mathrm{\bar{f}}_{bc}^{a}$ the complex structures of $sl(2,%
\mathbb{C})_{L}$ and $sl(2,\mathbb{C})_{R}.$ The commutators $\left[ A_{L%
\text{\textsc{b}}},A_{L\text{\textsc{c}}}\right] $ and $\left[ A_{R\text{%
\textsc{b}}},A_{R\text{\textsc{c}}}\right] $ show that traces $tr(\mathbf{A}%
_{L}\mathbf{A}_{L})$ and $tr(\mathbf{A}_{R}\mathbf{A}_{R})$ vanish
identically because $trJ_{a}^{L}=trJ_{a}^{R}=0$. For the case where $%
\boldsymbol{U}=\mathbf{A}_{L}$ and $\boldsymbol{V}=\mathbf{A}_{R}$, the
commutator $\left[ A_{L\text{\textsc{b}}},A_{R\text{\textsc{c}}}\right] $
and $tr(\mathbf{A}_{L}\mathbf{A}_{R})$ vanish due to the vanishing of $%
[J_{La},J_{Rb}].$\textrm{\ }

\begin{figure}[tbph]
\begin{center}
\includegraphics[width=7cm]{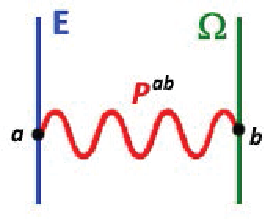}
\end{center}
\par
\vspace{-0.5cm}
\caption{Propagator $\left\langle E^{a}\left( \protect\zeta _{1}\right)
\Omega ^{b}\left( \protect\zeta _{2}\right) \right\rangle $ between a
vielbein line defects: $\mathbf{E}^{a}$ located at $\protect\zeta _{1}$ in $%
\mathbb{CP}^{1}$; and a spin connection line defect $\mathbf{\Omega }^{b}$
located at $\protect\zeta _{2}.$ The wavy line in red represents the
exchanged graviton state.}
\label{prop}
\end{figure}

\begin{figure}[tbph]
\begin{center}
\includegraphics[width=12cm]{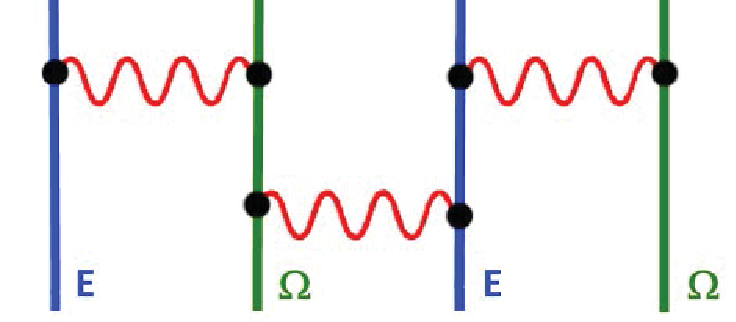}
\end{center}
\par
\vspace{-0.5cm}
\caption{Interaction between gravitational line defects. Two blue line
defects E interact through a green line defect $\Omega .$ Similarly, two
green line defects $\Omega $ couple through a blue line defect E. }
\label{02EW}
\end{figure}

Notice also that by using eq(\ref{EW}) and the property
\begin{equation}
\mathcal{P}_{\text{\textsc{ab}}L}^{ab}=+\mathcal{P}_{\text{\textsc{ab}}%
}^{ab}\qquad ,\qquad \mathcal{P}_{\text{\textsc{ab}}R}^{ab}=-\mathcal{P}_{%
\text{\textsc{ab}}}^{ab}
\end{equation}%
we can express the Chern-Simons two-points Green functions (\ref{3A}) in
terms of the two-points of the gravity fields. Straightforward calculations
lead, amongst others, to the following interesting constraint relations%
\begin{eqnarray}
\left\langle E_{\text{\textsc{a}}}^{a}\left( \zeta _{1}\right) \Omega _{%
\text{\textsc{b}}}^{b}\left( \zeta _{2}\right) \right\rangle +\left\langle
\Omega _{\text{\textsc{a}}}^{a}\left( \zeta _{1}\right) E_{\text{\textsc{b}}%
}^{b}\left( \zeta _{2}\right) \right\rangle  &=&\mathcal{P}_{\text{\textsc{ab%
}}}^{ab}\left( \zeta _{1}-\zeta _{2}\right)  \\
\left\langle E_{\text{\textsc{a}}}^{a}\left( \zeta _{1}\right) \Omega _{%
\text{\textsc{b}}}^{b}\left( \zeta _{2}\right) \right\rangle -\left\langle
\Omega _{\text{\textsc{a}}}^{a}\left( \zeta _{1}\right) E_{\text{\textsc{b}}%
}^{b}\left( \zeta _{2}\right) \right\rangle  &=&0
\end{eqnarray}%
As a result, the non trivial two-point Green functions of the gravity fields
$E_{\text{\textsc{a}}}^{a}\left( \zeta \right) $ and $\Omega _{\text{\textsc{%
b}}}^{b}\left( \zeta \right) $ are given by,%
\begin{eqnarray}
\left\langle E_{\text{\textsc{a}}}^{a}\left( \zeta _{1}\right) \Omega _{%
\text{\textsc{b}}}^{b}\left( \zeta _{2}\right) \right\rangle  &=&\frac{1}{2}%
\mathcal{P}_{\text{\textsc{ab}}}^{ab}\left( \zeta _{1}-\zeta _{2}\right)
\label{p1} \\
\left\langle \Omega _{\text{\textsc{a}}}^{a}\left( \zeta _{1}\right) \Omega
_{\text{\textsc{b}}}^{b}\left( \zeta _{2}\right) \right\rangle  &=&0
\label{p2} \\
\left\langle E_{\text{\textsc{a}}}^{a}\left( \zeta _{1}\right) E_{\text{%
\textsc{b}}}^{b}\left( \zeta _{2}\right) \right\rangle  &=&0  \label{p3}
\end{eqnarray}%
A\textrm{\ graphical description of the above propagator is illustrated in}
the Figure \textbf{\ref{prop} }where a vielbein line defect $\boldsymbol{E}%
^{a}$ at $\zeta _{1}$ exchanges a topological graviton with a spin
connection line defect $\boldsymbol{\Omega }^{a}$ at $\zeta _{2}$.
Using the 1-forms $\boldsymbol{E}^{a}=d\zeta ^{\text{\textsc{a}}}E_{\text{%
\textsc{a}}}^{a}$\textrm{\ }and\textrm{\ }$\boldsymbol{\Omega }^{a}=d\zeta ^{%
\text{\textsc{a}}}\Omega _{\text{\textsc{a}}}^{a}$ as well as the 2-form,%
\begin{equation}
\mathcal{P}^{ab}\left( \zeta \right) =\frac{1}{2}d\zeta ^{\text{\textsc{a}}%
}\wedge d\zeta ^{\text{\textsc{b}}}\mathcal{P}_{\text{\textsc{ab}}%
}^{ab}\left( \zeta \right)
\end{equation}%
we can present the gravitational two-points Green functions (\ref{p1}-\ref%
{p3}) in a shortened form as follows%
\begin{equation}
\left\langle \boldsymbol{E}^{a}\left( \zeta _{1}\right) \boldsymbol{\Omega }%
^{b}\left( \zeta _{2}\right) \right\rangle =\mathcal{P}^{ab}\left( \zeta
_{1}-\zeta _{2}\right)   \label{pp}
\end{equation}%
and%
\begin{equation}
\left\langle \boldsymbol{E}^{a}\left( \zeta _{1}\right) \boldsymbol{E}%
^{b}\left( \zeta _{2}\right) \right\rangle =\left\langle \boldsymbol{\Omega }%
^{a}\left( \zeta _{1}\right) \boldsymbol{\Omega }^{b}\left( \zeta
_{2}\right) \right\rangle =0
\end{equation}%
where $\mathcal{P}^{ab}\left( \zeta \right) =\delta ^{ab}\mathcal{P}\left(
\zeta \right) $ with scalar 2-form,%
\begin{equation}
\mathcal{P}\left( \zeta \right) =\frac{1}{2\pi \left( x^{2}+y^{2}+\zeta \bar{%
\zeta}\right) ^{2}}\left( xdy\wedge d\zeta +yd\bar{\zeta}\wedge dx+2\zeta
dx\wedge dy\right)   \label{pq}
\end{equation}%
\textrm{T}he vanishing propagators $\left\langle \boldsymbol{E}\left( \zeta
_{1}\right) \boldsymbol{E}\left( \zeta _{2}\right) \right\rangle $ and $%
\left\langle \boldsymbol{\Omega }\left( \zeta _{1}\right) \boldsymbol{\Omega
}\left( \zeta _{2}\right) \right\rangle $ given by (\ref{p2}-\ref{p3}) show
that \textrm{identical} topological gravitational line defects cannot
interact directly; they \textrm{must} couple either indirectly \textrm{via a
different topological defect} as exhibited by the Figure \textbf{\ref{02EW},}
or via quantum corrections.

\subsection{Line defects in 4D topological gravity}

\qquad Because of the flatness of the gauge curvatures (\ref{f2l}-\ref{f2r}%
), no gauge invariant observable can be built out of the gauge 2-forms $%
\boldsymbol{F}_{2}^{L}$ and $\boldsymbol{F}_{2}^{R}$\ as they vanish on
shell. Observables in this 4D topological gravity are given by surface and
line defects \cite{1BC,1BD} sitting in $\mathbb{R}^{2}\times \mathbb{%
CP}^{1}$; they are obtained by extending results from the CWY theory while
using the loop matrices
\begin{equation}
\Phi _{\boldsymbol{A}_{L}}\left[ \mathrm{\gamma }_{z}\right]
=\doint\nolimits_{\mathrm{\gamma }_{z}}\boldsymbol{A}_{L}\qquad ,\qquad \Phi
_{\boldsymbol{A}_{R}}\left[ \mathrm{\gamma }_{z}\right] =\doint\nolimits_{%
\mathrm{\gamma }_{z}}\boldsymbol{A}_{R}  \label{g1}
\end{equation}%
with $\Phi _{\boldsymbol{A}_{L}/\boldsymbol{A}_{R}}$ valued in $sl(2,\mathbb{%
C})_{L/R}.$

\subsubsection{Gravitational holonomies}

The above loop matrices are gauge holonomies of the topological CS fields $%
\boldsymbol{A}_{L}$ and $\boldsymbol{A}_{R}.$With the help of the change (%
\ref{EW}), we can also express these gravitational holonomies of 4D gravity
in terms of the gravitational 1-form potentials as follows%
\begin{equation}
\Phi _{\Omega }\left[ \mathrm{\gamma }_{z}\right] =\doint\nolimits_{\mathrm{%
\gamma }_{z}}\boldsymbol{\Omega }\qquad ,\qquad \Phi _{E}\left[ \mathrm{%
\gamma }_{w}\right] =\doint\nolimits_{\mathrm{\gamma }_{w}}\boldsymbol{E}
\label{g2}
\end{equation}%
These novel loop matrices are valued in $sl(2,\mathbb{C})$ given by eq(\ref%
{dia}) and generated by the diagonal $J_{a}$s; they define the gravitational
holonomies in opposition to the gauge holonomies (\ref{g1}). Notice that as
in eq(\ref{g1}), the real lines $\mathrm{\gamma }_{z}$ and $\mathrm{\gamma }%
_{w}$ appearing (\ref{g2}) are loops spreading in $\mathbb{R}^{2}$; but
sitting at the point z and w in the complex projective line $\mathbb{CP}^{1}$%
. These loop matrices\thinspace are given by eq(\ref{fl}); because $\mathrm{%
\gamma }_{z}$ and $\mathrm{\gamma }_{w}$ spread in the topological plane $%
\mathbb{R}^{2}$, they read explicitly as follows%
\begin{eqnarray}
\Phi _{E}{\small [}\mathrm{\gamma }_{z}{\small ]} &=&\doint\nolimits_{%
\mathrm{\gamma }_{z}}\left( \mathbf{E}_{x}dx+\mathbf{E}_{y}dy\right) \\
\Phi _{\Omega }{\small [}\mathrm{\gamma }_{w}{\small ]} &=&\doint\nolimits_{%
\mathrm{\gamma }_{w}}\left( \mathbf{\Omega }_{x}dx+\mathbf{\Omega }%
_{y}dy\right)
\end{eqnarray}%
Using $\boldsymbol{A}_{L}=J_{a}^{L}A_{L}^{a}$ and $\boldsymbol{A}%
_{R}=J_{a}^{R}A_{R}^{a}$ as well as $\mathbf{E}=J_{a}E^{a}$ and $\mathbf{%
\Omega }=J_{a}\Omega ^{a},$ the gauge and gravity loop matrices expand like
\begin{equation}
\begin{tabular}{lll}
$\Phi _{\boldsymbol{A}_{L}}\left[ \mathrm{\gamma }_{z}\right] $ & $=$ & $%
J_{a}^{L}\Phi _{\boldsymbol{A}_{L}}^{a}{\small [}\mathrm{\gamma }_{z}{\small %
]}$ \\
$\Phi _{\boldsymbol{A}_{R}}\left[ \mathrm{\gamma }_{z}\right] $ & $=$ & $%
J_{a}^{R}\Phi _{\boldsymbol{A}_{R}}^{a}{\small [}\mathrm{\gamma }_{z}{\small %
]}$ \\
$\Phi _{E}{\small [}\mathrm{\gamma }_{z}{\small ]}$ & $=$ & $J_{a}\Phi
_{E}^{a}{\small [}\mathrm{\gamma }_{z}{\small ]}$ \\
$\Phi _{\Omega }{\small [}\mathrm{\gamma }_{z}{\small ]}$ & $=$ & $J_{a}\Phi
_{\Omega }^{a}{\small [}\mathrm{\gamma }_{z}{\small ]}$%
\end{tabular}
\label{AE}
\end{equation}%
with components as%
\begin{equation}
\begin{tabular}{lll}
$\Phi _{\boldsymbol{A}_{L}}^{a}\left[ \mathrm{\gamma }_{z}\right] $ & $=$ & $%
\doint\nolimits_{\mathrm{\gamma }_{z}}\boldsymbol{A}_{L}^{a}$ \\
$\Phi _{\boldsymbol{A}_{R}}^{a}\left[ \mathrm{\gamma }_{z}\right] $ & $=$ & $%
\doint\nolimits_{\mathrm{\gamma }_{z}}\boldsymbol{A}_{R}^{a}$ \\
$\Phi _{\Omega }^{a}\left[ \mathrm{\gamma }_{z}\right] $ & $=$ & $%
\doint\nolimits_{\mathrm{\gamma }_{z}}\boldsymbol{\Omega }^{a}$ \\
$\Phi _{E}^{a}\left[ \mathrm{\gamma }_{z}\right] $ & $=$ & $\doint\nolimits_{%
\mathrm{\gamma }_{z}}\boldsymbol{E}^{a}$%
\end{tabular}%
\end{equation}
These holonomy components follow from the projections%
\begin{equation}
\begin{tabular}{lll}
$\Phi _{\boldsymbol{A}_{L/R}}^{a}$ & $=$ & $q_{L/R}^{ab}tr(J_{b}^{L/R}\Phi _{%
\boldsymbol{A}_{L/R}})$ \\
$\Phi _{\boldsymbol{\Omega }}^{a}$ & $=$ & $q^{ab}tr(J_{b}\Phi _{\boldsymbol{%
\Omega }})$ \\
$\Phi _{\boldsymbol{E}}^{a}$ & $=$ & $q^{ab}tr(J_{b}\Phi _{\boldsymbol{E}})$%
\end{tabular}%
\end{equation}%
with $q_{ab}^{L/R}=tr(J_{a}^{L/R}J_{b}^{L/R})$ and $q_{L/R}^{ab}$ its
inverse $q_{L/R}^{ba}$. The gravitational loop matrices (\ref{AE}) satisfy
the commutation property%
\begin{equation}
\Phi _{\boldsymbol{A}_{L}}{\small [}\mathrm{\gamma }_{z}{\small ]}\bullet
\Phi _{\boldsymbol{A}_{R}}{\small [}\mathrm{\gamma }_{z}{\small ]}=\Phi _{%
\boldsymbol{A}_{R}}{\small [}\mathrm{\gamma }_{z}{\small ]}\bullet \Phi _{%
\boldsymbol{A}_{L}}{\small [}\mathrm{\gamma }_{z}{\small ]}
\end{equation}%
as well as the non commutation relations%
\begin{equation}
\Phi _{\Omega }{\small [}\mathrm{\gamma }_{z}{\small ]}\bullet \Phi _{E}%
{\small [}\mathrm{\gamma }_{z}{\small ]}-\Phi _{E}{\small [}\mathrm{\gamma }%
_{z}{\small ]}\bullet \Phi _{\Omega }{\small [}\mathrm{\gamma }_{z}{\small ]}%
=\Psi _{\Omega E}{\small [}\mathrm{\gamma }_{z}{\small ]}  \label{aew}
\end{equation}%
with%
\begin{equation}
\Psi _{\Omega E}{\small [}\mathrm{\gamma }_{z}{\small ]}=J_{a}\Psi _{\Omega
E}^{a}{\small [}\mathrm{\gamma }_{z}{\small ]}
\end{equation}%
and%
\begin{equation}
(\Psi _{\Omega E}{\small [}\mathrm{\gamma }_{z}{\small ])}_{a}=\varepsilon
_{abc}\Phi _{\Omega }^{b}{\small [}\mathrm{\gamma }_{z}{\small ]}\Phi
_{E}^{c}{\small [}\mathrm{\gamma }_{z}{\small ]}
\end{equation}%
By using eq(\ref{EW}), we have the relationships%
\begin{equation}
\begin{tabular}{lll}
$\Phi _{\boldsymbol{A}_{L}}^{a}\left[ \mathrm{\gamma }_{z}\right] $ & $=$ & $%
\Phi _{\Omega }^{a}\left[ \mathrm{\gamma }_{z}\right] +\Phi _{\boldsymbol{E}%
}^{a}\left[ \mathrm{\gamma }_{z}\right] $ \\
$\Phi _{\boldsymbol{A}_{R}}^{a}\left[ \mathrm{\gamma }_{z}\right] $ & $=$ & $%
\Phi _{\Omega }^{a}\left[ \mathrm{\gamma }_{z}\right] -\Phi _{\boldsymbol{E}%
}^{a}\left[ \mathrm{\gamma }_{z}\right] $ \\
$\Phi _{\Omega }^{a}\left[ \mathrm{\gamma }_{z}\right] $ & $=$ & $\frac{1}{2}%
\left( \Phi _{\boldsymbol{A}_{L}}^{a}\left[ \mathrm{\gamma }_{z}\right]
+\Phi _{\boldsymbol{A}_{R}}^{a}\left[ \mathrm{\gamma }_{z}\right] \right) $
\\
$\Phi _{\boldsymbol{E}}^{a}\left[ \mathrm{\gamma }_{z}\right] $ & $=$ & $%
\frac{1}{2}\left( \Phi _{\boldsymbol{A}_{L}}^{a}\left[ \mathrm{\gamma }_{z}%
\right] -\Phi _{\boldsymbol{A}_{R}}^{a}\left[ \mathrm{\gamma }_{z}\right]
\right) $%
\end{tabular}%
\end{equation}%
leading to
\begin{eqnarray}
\Phi _{\Omega }^{a}\left[ \mathrm{\gamma }_{z}\right] &=&\frac{q^{ab}}{2}%
\left[ tr(J_{b}^{L}\Phi _{\boldsymbol{A}_{L}}\left[ \mathrm{\gamma }_{z}%
\right] )+tr(J_{b}^{R}\Phi _{\boldsymbol{A}_{R}}\left[ \mathrm{\gamma }_{z}%
\right] )\right] \\
\Phi _{E}^{a}\left[ \mathrm{\gamma }_{z}\right] &=&\frac{q^{ab}}{2}\left[
tr(J_{b}^{L}\Phi _{\boldsymbol{A}_{L}}\left[ \mathrm{\gamma }_{z}\right]
)-tr(J_{b}^{R}\Phi _{\boldsymbol{A}_{R}}\left[ \mathrm{\gamma }_{z}\right] )%
\right]
\end{eqnarray}%
where we have used $q_{L}^{ab}=q_{R}^{ab}=q^{ab}$. By setting%
\begin{equation}
tr(J_{a}J_{b}^{L})=\chi _{ab}^{L}\qquad ,\qquad tr(J_{a}J_{b}^{R})=\chi
_{ab}^{R}
\end{equation}%
they read as follows%
\begin{eqnarray}
\Phi _{\Omega }^{a}\left[ \mathrm{\gamma }_{z}\right] &=&\frac{q^{ab}}{2}%
\left( \chi _{bc}^{L}\Phi _{\boldsymbol{A}_{L}}^{c}\left[ \mathrm{\gamma }%
_{z}\right] +\chi _{bc}^{R}\Phi _{\boldsymbol{A}_{R}}^{c}\left[ \mathrm{%
\gamma }_{z}\right] \right) \\
\Phi _{E}^{a}\left[ \mathrm{\gamma }_{z}\right] &=&\frac{q^{ab}}{2}\left(
\chi _{bc}^{L}\Phi _{\boldsymbol{A}_{L}}^{c}\left[ \mathrm{\gamma }_{z}%
\right] -\chi _{bc}^{R}\Phi _{\boldsymbol{A}_{R}}^{c}\left[ \mathrm{\gamma }%
_{z}\right] \right)
\end{eqnarray}%
By using $\chi _{bc}^{L}=\chi _{bc}^{R}=q_{bc}$, we bring the above
relations to a simpler form%
\begin{eqnarray}
\Phi _{\Omega }^{a}\left[ \mathrm{\gamma }_{z}\right] &=&\frac{1}{2}\left(
\Phi _{\boldsymbol{A}_{L}}^{a}\left[ \mathrm{\gamma }_{z}\right] +\Phi _{%
\boldsymbol{A}_{R}}^{a}\left[ \mathrm{\gamma }_{z}\right] \right) \\
\Phi _{E}^{a}\left[ \mathrm{\gamma }_{z}\right] &=&\frac{1}{2}\left( \Phi _{%
\boldsymbol{A}_{L}}^{a}\left[ \mathrm{\gamma }_{z}\right] -\Phi _{%
\boldsymbol{A}_{R}}^{a}\left[ \mathrm{\gamma }_{z}\right] \right)
\end{eqnarray}%
showing that gravitational holonomies $\Phi _{\Omega }^{a}\left[ \mathrm{%
\gamma }_{z}\right] $ and $\Phi _{E}^{a}\left[ \mathrm{\gamma }_{z}\right] $
are given by linear combinations of the left/right gauge holonomies $\Phi _{%
\boldsymbol{A}_{L}}^{a}$ and $\Phi _{\boldsymbol{A}_{R}}^{a}$ used in eq(\ref%
{W}). The $\Phi _{\Omega }^{a}\left[ \mathrm{\gamma }_{z}\right] $ is the
mean value of $\Phi _{\boldsymbol{A}_{L}}^{a}\left[ \mathrm{\gamma }_{z}%
\right] $ and $\Phi _{\boldsymbol{A}_{R}}^{a}\left[ \mathrm{\gamma }_{z}%
\right] $ while the $\Phi _{E}^{a}\left[ \mathrm{\gamma }_{z}\right] $ is
the relative ---reduced--- value.

Moreover using line defects of the CWY theory such as the Wilson loops (\ref%
{W}), we see that due to eqs(\ref{4dg}-\ref{r}), they appear in \textrm{both}
varieties left $W_{\boldsymbol{A}_{L}}\left[ \mathrm{\gamma }_{z}\right] $
and right $W_{\boldsymbol{A}_{R}}\left[ \mathrm{\gamma }_{z}\right] $ given
by%
\begin{eqnarray}
W_{\boldsymbol{A}_{L}}\left[ \mathrm{\gamma }_{z}\right] &=&Pe^{\Phi _{A_{L}}%
\left[ \mathrm{\gamma }_{z}\right] } \\
W_{\boldsymbol{A}_{R}}\left[ \mathrm{\gamma }_{z}\right] &=&Pe^{\Phi _{A_{R}}%
\left[ \mathrm{\gamma }_{z}\right] }
\end{eqnarray}%
By using eq(\ref{EW}), these topological gauge lines can be presented in
terms of gravitational defects $W_{E}\left[ \mathrm{\gamma }_{z}\right] $
and $W_{\Omega }\left[ \mathrm{\gamma }_{z}\right] $ like%
\begin{eqnarray}
W_{\boldsymbol{A}_{L}}\left[ \mathrm{\gamma }_{z}\right] &=&P\exp \left(
\doint\nolimits_{\mathrm{\gamma }_{z}}\Omega +\doint\nolimits_{\mathrm{%
\gamma }_{z}}E\right)  \label{l1} \\
W_{\boldsymbol{A}_{R}}\left[ \mathrm{\gamma }_{w}\right] &=&P\exp \left(
\doint\nolimits_{\mathrm{\gamma }_{z}}\Omega -\doint\nolimits_{\mathrm{%
\gamma }_{z}}E\right)  \label{l2}
\end{eqnarray}%
These relations show that in the 4D topological gravity (\ref{4dg}), one
distinguishes two non commuting gravitational Wilson-like lines $W_{E}\left[
\mathrm{\gamma }_{z}\right] $ and $W_{\Omega }\left[ \mathrm{\gamma }_{z}%
\right] $ defined as%
\begin{eqnarray}
W_{E}\left[ \mathrm{\gamma }_{z}\right] &=&P\exp \left( \doint\nolimits_{%
\mathrm{\gamma }_{z}}\mathbf{E}\right)  \label{1g} \\
W_{\Omega }\left[ \mathrm{\gamma }_{z}\right] &=&P\exp \left(
\doint\nolimits_{\mathrm{\gamma }_{z}}\mathbf{\Omega }\right)  \label{11g}
\end{eqnarray}%
These gravitational line defects give basic topological observables that
constitute building blocks towards the study of quantum integrability in the
obtained 4D topological gravity.
\begin{figure}[tbph]
\begin{center}
\includegraphics[width=12cm]{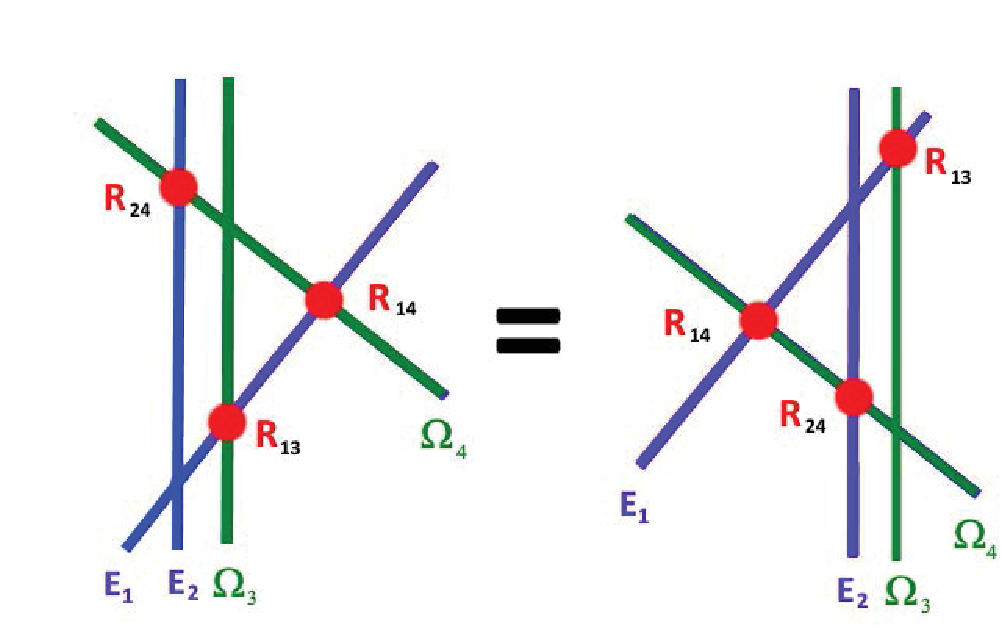}
\end{center}
\par
\vspace{-0.5cm}
\caption{Yang-Baxter equations for the gravitational line defects $\left(
\mathbf{E}_{1},\mathbf{E}_{2}\right) $ and $\left( \mathbf{\Omega }_{1},%
\mathbf{\Omega }_{2}\right) $: \ 2+2 crossing gravitational line defects.
Because of eq(\protect\ref{p1}-\protect\ref{p3}), the contributions are
given by the crossing of vielbein lines with spin connection lines (red
points).}
\label{RRR}
\end{figure}
Other basic building blocks will be given below.

\subsubsection{Gravitational integrable equations}

Applications of the gravitational line defects constructed above to quantum
integrable systems and brane realisations in type II strings along the line
of \textrm{\cite{1DA}} will be given in a future occasion. Below, we give
partial results by considering below two particular systems of crossing line
defects.

\ \ \ \

\begin{figure}[tbph]
\begin{center}
\includegraphics[width=10cm]{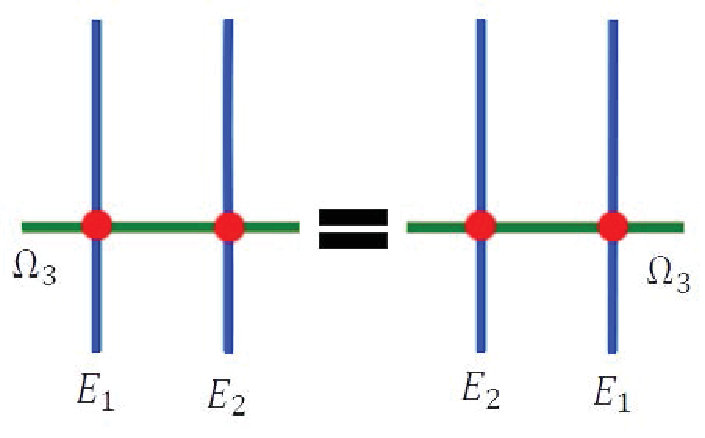}
\end{center}
\par
\vspace{-0.5cm}
\caption{The crossing of three line defects with one having a different
color.}
\label{A}
\end{figure}

\textbf{A)} \emph{gravitational Yang-Baxter like system}\newline
Here, we give two examples of integrable relations having a description in
terms of crossing gravitational line defects (\ref{1g}-\ref{11g}). The first
example we give concerns the gravitational Yang-Baxter like equation; it is
represented graphically by the Figure\emph{\ }\textbf{\ref{RRR}}. It is
given by a set of 2+2 non trivial crossings \textrm{of} topological line
defects involving:

\begin{description}
\item[$\left( \mathbf{a}\right) $] two Wilson line defects of vielbein like W%
$_{E}$[$\mathrm{\gamma }_{z_{1}}{\small ]}$ and W$_{E}$[$\mathrm{\gamma }%
_{z_{2}}{\small ]}$; they are given by the lines $\mathrm{\gamma }_{z_{1}}$
and $\mathrm{\gamma }_{z_{2}}$\ spreading in the topological plane $\mathbb{R%
}^{2}$; and respectively located at $z_{1}$ and $z_{2}$ in the holomorphic $%
\mathbb{CP}^{1}{\small .}$

\item[$\left( \mathbf{b}\right) $] two Wilson line defects of spin
connection like W$_{\Omega }$[$\mathrm{\xi }_{w_{3}}{\small ]},$ W$_{\Omega
} $[$\mathrm{\xi }_{w_{4}}{\small ]}$; they are given by the lines $\mathrm{%
\xi }_{w_{3}}$ and $\mathrm{\xi }_{w_{4}}$\ spreading in $\mathbb{R}^{2}$
and located at $w_{3}$ and $w_{4}$.
\end{description}
The positions of these gravitational topological lines in the plane $\mathbb{%
R}^{2}$ are such that they have crossings as depicted by the Figure \textbf{%
\ref{02EW}}. This particular configuration of four lines have three
intersection points given by%
\begin{equation}
\begin{tabular}{lll}
$P_{13}$ & $=$ & $\mathrm{\gamma }_{z_{1}}\cap \mathrm{\xi }_{w_{3}}$ \\
$P_{14}$ & $=$ & $\mathrm{\gamma }_{z_{1}}\cap \mathrm{\xi }_{w_{4}}$ \\
$P_{23}$ & $=$ & $\mathrm{\gamma }_{z_{2}}\cap \mathrm{\xi }_{w_{3}}$%
\end{tabular}%
\end{equation}%
but no intersection point $P_{24}$ because W$_{E}$[$\mathrm{\gamma }_{z_{2}}%
{\small ]}$ and W$_{\Omega }$[$\mathrm{\xi }_{w_{3}}{\small ]}$ are
parallel. Then, we use the topological symmetry in the topological $\mathbb{R%
}^{2}$ to move the two vertical line defects W$_{E}$[$\mathrm{\gamma }%
_{z_{2}}{\small ]}$ and W$_{\Omega }$[$\mathrm{\xi }_{w_{3}}{\small ]}$ from
the left side of the node $\mathbf{R}_{14}$ to the right side. This shifting
process in $\mathbb{R}^{2}$ generates the integrability equations%
\begin{equation}
\mathbf{R}_{13}\otimes \mathbf{R}_{14}\otimes \mathbf{R}_{24}=\mathbf{R}%
_{24}\otimes \mathbf{R}_{14}\otimes \mathbf{R}_{13}  \label{ybe}
\end{equation}%
with $z_{ij}=z_{i}-z_{j}$; and where $\mathbf{R}_{ij}=\mathbf{R}\left(
z_{ij}\right) $ is the Yang-Baxter R-matrix of integrable quantum systems
\cite{1A} acting in the tensor space
\begin{equation}
V_{1}\otimes V_{2}\otimes V_{3}\otimes V_{4}
\end{equation}%
A semi classical expression of $\mathbf{R}_{ij}$ is given by
\begin{equation}
\mathbf{R}_{\text{\textsc{ab}}}^{\text{\textsc{cd}}}\left(
z_{i}-z_{j}\right) =\delta _{\text{\textsc{a}}}^{\text{\textsc{c}}}\delta _{%
\text{\textsc{b}}}^{\text{\textsc{d}}}+\frac{\hbar }{z_{i}-z_{j}}\mathbf{c}_{%
\text{\textsc{ab}}}^{\text{\textsc{cd}}}+O\left( \hbar ^{2}\right)
\end{equation}%
where $\mathbf{c}_{\text{\textsc{ab}}}^{\text{\textsc{cd}}}$ stands for the
double Casimir of the gauge symmetry; its value for $sl\left( N,\mathbb{C}%
\right) $ is $\delta _{\text{\textsc{a}}}^{\text{\textsc{c}}}\delta _{\text{%
\textsc{b}}}^{\text{\textsc{d}}}.$ The second example we give involves the
crossings of three topological lines; this system can be: $\left( i\right) $
completely trivial as for the cases where the three lines are of same
nature; that is having the same color; or $\left( ii\right) $ lead to
commutative products as for the situation depicted by the the Figure \textbf{%
\ref{A}}.
\textrm{\ }For this line system, the obtained integrability equation leads
the following commutation relation resulting from the move of the line E1
from the left to the right of the line E2,%
\begin{equation}
\mathbf{R}_{13}\otimes \mathbf{R}_{23}=\mathbf{R}_{23}\otimes \mathbf{R}_{13}
\end{equation}

\ \ \

\textbf{B) }\emph{gravitational RLL- like system}\newline
This is an interesting system of topological 4D gravity generalising the
usual RLL relations in the CWY gauge theory depicted by the Figure \textbf{%
\ref{M1}}-(b). It involves a set of 2+2 non trivial crossings of topological
line defects as described here below:

\begin{description}
\item[$\left( \mathbf{\protect\alpha }\right) $] two Wilson-like lines W$%
_{E} $[$\mathrm{\gamma }_{z_{1}}{\small ]},$ W$_{\Omega _{2}}$[$\mathrm{\xi }%
_{w_{2}}{\small ]}$ with quantum states respectively characterised by
highest weights (HW) $\boldsymbol{e}$ and $\boldsymbol{\lambda }$ of the
gauge symmetry ---here SL(2,$\mathbb{C}$)---. These two HWs can be taken
equal.

\item[$\left( \mathbf{\protect\beta }\right) $] two dual line defects
described by 't Hooft lines tH$_{E}$[$\mathrm{\gamma }_{z_{3}}{\small ]}$
and tH$_{\Omega }$[$\mathrm{\xi }_{w_{4}}{\small ]}$.
\begin{figure}[tbph]
\begin{center}
\includegraphics[width=15cm]{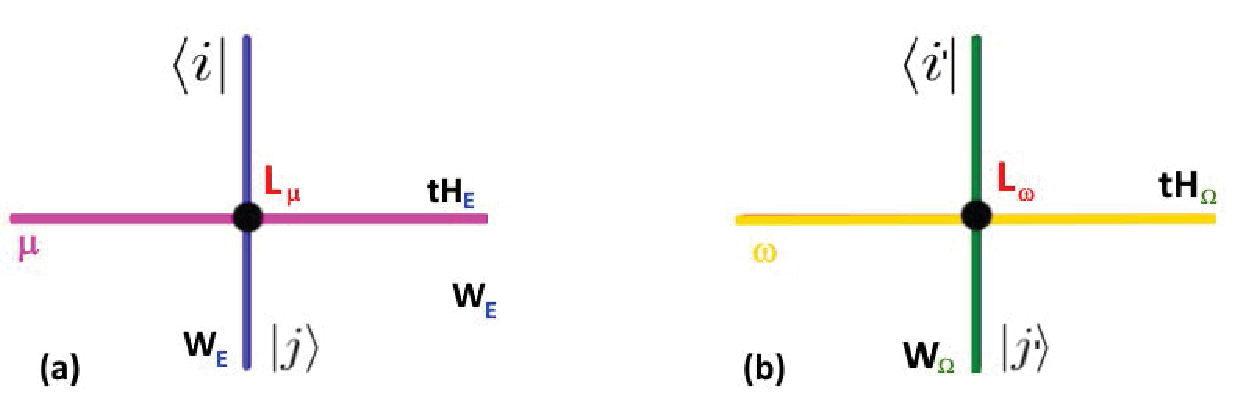}
\end{center}
\par
\vspace{-0.5cm}
\caption{'t Hooft lines tH$_{\protect\mu }$ and tH$_{\protect\omega }$
respectively dual to the Wilson lines W$_{E}$ and W$_{\Omega }$. On the left
the vielbein pair given by a horizontal magenta 't Hooft line dual to
vertical blue Wilson line line \ On the right the spin connection pair given
by a horizontal yellow t Hooft line dual to vertical green Wilson line.}
\label{th}
\end{figure}
These 't Hooft lines are characterised by coweights $\mu $ and $\omega $
respectively dual to the weights $\boldsymbol{e}$ and $\boldsymbol{\lambda }$%
. These 't Hooft lines are depicted by the red and the Yellow lines in the
Figure \textbf{\ref{th}}.
\end{description}

\begin{figure}[tbph]
\begin{center}
\includegraphics[width=14cm]{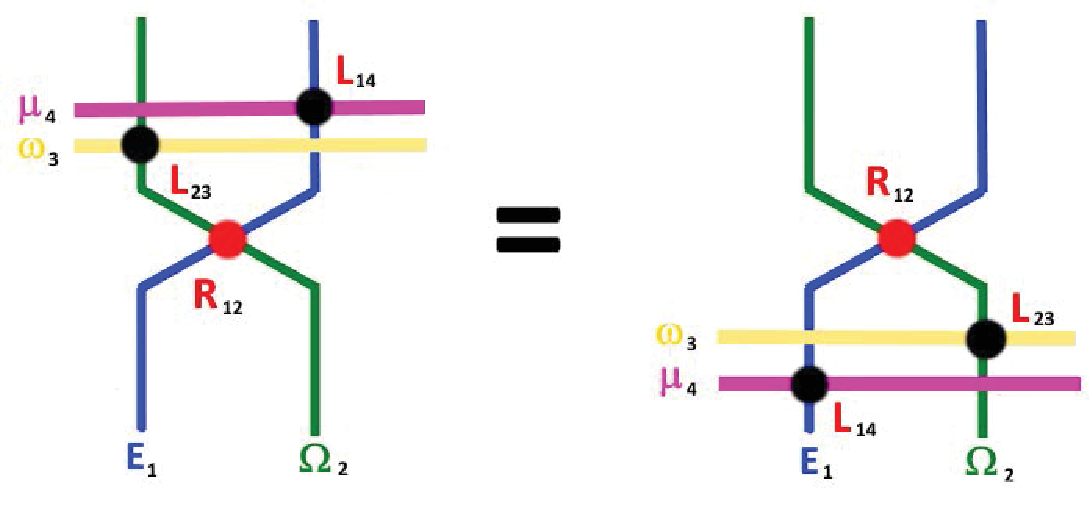}
\end{center}
\par
\vspace{-0.5cm}
\caption{An interacting system of 2+2 lines defects made of : $\left(
i\right) $ a blue vielbein line defect $W_{E}\left[ \mathrm{\protect\gamma }%
_{z_{1}}\right] $ crossed by spin line $W_{\Omega }\left[ \mathrm{\protect%
\xi }_{w_{1}}\right] $. $\left( ii\right) $ two horizontal 't Hooft lines tH$%
_{E}\left[ \mathrm{\protect\gamma }_{z_{3}}\right] $ and $tH_{\Omega }\left[
\mathrm{\protect\xi }_{w_{2}}\right] $ traversing the Wilson lines. The
identification of the two ways gives the gravitational RLL-like equation.}
\label{RTT}
\end{figure}

Using the gravitational Wilson-like lines W$_{E}$[$\mathrm{\gamma }_{z_{1}}%
{\small ]},$ W$_{\Omega _{2}}$[$\mathrm{\xi }_{w_{2}}{\small ]}$ and their
magnetic duals tH$_{E}$[$\mathrm{\gamma }_{z_{3}}{\small ]},$ tH$_{\Omega }$[%
$\mathrm{\xi }_{w_{4}}{\small ]}$ all of them spreading in the topological
plane with intersections as depicted by the Figure \textbf{\ref{RTT},}
one can derive the integrability equation describing the solvability of this
gravitational system. Indeed, starting from the left side in the Figure
\textbf{\ref{RTT}} with horizontal 't Hooft lines (in yellow and magenta
colors); then moving these two horizontal lines below the vertex $\mathbf{R}%
_{12}$, we obtain the gravitational RLL- like in topological 4D gravity. It
reads formally as
\begin{equation}
\mathbf{R}_{12}\otimes \mathbf{L}_{23}\otimes \mathbf{L}_{14}=\mathbf{L}%
_{14}\otimes \mathbf{L}_{23}\otimes \mathbf{R}_{12}  \label{r14}
\end{equation}
where $\mathbf{L}_{14}$ and $\mathbf{L}_{23}$ are Lax operators whose
expressions can be obtained by using the formula eq(\ref{lm}), for technical
computations of the $\mathbf{L}_{ij}$'s in CWY theory see \textrm{\cite{1DB}}. The explicit expression of the integrability relation (\ref{r14})
reads as follows%
\begin{equation}
\mathbf{R}_{\text{\textsc{ef}}}^{\text{\textsc{ad}}}\left(
z_{1}-z_{2}\right) \mathbf{L}_{\text{\textsc{b}}}^{\text{\textsc{e}}}\left(
z_{1}\right) \mathbf{L}_{\text{\textsc{c}}}^{\text{\textsc{f}}}\left(
z_{2}\right) =\mathbf{L}_{\text{\textsc{e}}}^{\text{\textsc{a}}}\left(
z_{2}\right) \mathbf{L}_{\text{\textsc{f}}}^{\text{\textsc{d}}}\left(
z_{1}\right) \mathbf{R}_{\text{\textsc{ad}}}^{\text{\textsc{ef}}}\left(
z_{1}-z_{2}\right)
\end{equation}

\section{Conclusion and comments}

\qquad In this study, we have constructed a novel integrable 4D topological
gravity obtained by extending the CWY method, used in the derivation of the
so-called 4D Chern-Simons of integrable spin chains, to the 3D Anti de
Sitter gravity. Here, the CWY method has been applied to the AdS$_{3}$
gravity; thus leading to an emergent 4D topological gravity with observables
given by topological gravitational defects. Concretely, the field action $%
\mathcal{S}_{4D}^{grav}$ of the obtained 4D gravity is given by eqs(\ref{4dg}%
\textbf{-}\ref{r}) or equivalently by (\ref{419});%
\begin{equation}
\mathcal{S}_{4D}^{gravity}=\dint\nolimits_{R^{2}\times \mathbb{CP}%
^{1}}dz\wedge E^{a}\wedge \left( d\Omega _{a}+\frac{1}{2}\varepsilon
_{abc}\Omega ^{b}\wedge \Omega ^{c}+\xi \varepsilon _{abc}E^{b}\wedge
E^{c}\right)  \label{4sd}
\end{equation}%
It has been obtained by using the Chern-Simons formulation of AdS$_{3}$
gravity as shown by eqs (\ref{1}) and (\ref{3gr}). Moreover, because $%
\mathcal{S}_{AdS_{3}}^{grav}$ is formulated in terms of two copies of
Chern-Simons fields $A_{L}$ and $A_{R}$, we ended up with two particular
gravitational lines defects $W_{E}\left[ \mathrm{\gamma }_{z}\right] $ and $%
W_{\Omega }\left[ \mathrm{\gamma }_{z}\right] $ respectively termed as
vielbein and spin connection lines. These line defects are given by eqs(\ref%
{l1}-\ref{11g}) and were used to derive partial results regarding their
crossings. Applications of this construction in quantum integrability and
embeddings in string theory as well as in link with BTZ black-hole will be
considered in a future occasion.

In the end of this investigation, we want to make a comment \textrm{%
regarding the extension of the} AdS$_{3}$/CFT$_{2}$ correspondence. Applying
the AdS/CFT duality to the CWY theory, one expects to have an emergent 4D
bulk/3D edge correspondence for topological 4D gravity. Below, we give an
argument indicating that this emergent 4D bulk/3D edge constitute an
interesting example of Gravity/Gauge duality where $\left( i\right) $ the
gravity sector is given by the obtained topological 4D gravity described by
the field action $\mathcal{S}_{4D}^{gravity}$ (\ref{4sd}); and $\left(
ii\right) $ the gauge sector is given by the topological 3D Chern-Simons
with field action $\mathcal{S}_{3D}^{{\small CS}}$ as in (\ref{3d}). Indeed,
in this conjectured 4D/3D correspondence, gravitational line defects $%
\mathrm{\gamma }_{z}$ spreading in the $\mathbb{R}^{2}$ plane of the 4D
space $\boldsymbol{M}_{4}=\mathbb{R}^{2}\times \mathbb{CP}^{1}$ gets mapped
into point-like particles on the 3D boundary
\begin{equation}
\partial \boldsymbol{M}_{4}=\left( \partial \mathbb{R}^{2}\right) \times
\mathbb{CP}^{1}\qquad ,\qquad \partial \left( \mathbb{CP}^{1}\right)
=\emptyset
\end{equation}%
Because $\mathcal{S}_{4D}^{gravity}$ on $\boldsymbol{M}_{4}$ is topological,
the dual gauge theory on $\partial \boldsymbol{M}_{4}=\boldsymbol{M}_{3}$
should be topological; thus justifying the $\mathcal{S}_{3D}^{{\small CS}}$
on the 3D boundary. This 4D/3D duality can be made more explicit by thinking
about the topological plane $\mathbb{R}^{2}$ in terms of the fibration $%
\mathbb{R}_{>0}\times \mathbb{S}^{1}$ with $\left( \partial \mathbb{R}%
^{2}\right) \sim \mathbb{S}^{1}$; then the 3D boundary $\partial \boldsymbol{%
M}_{4}$ is given by $\mathbb{S}^{1}\times \mathbb{S}^{2}$ which is
isomorphic to a 3-sphere $\mathbb{S}^{3}.$ Within this view, we see that%
\begin{equation}
\mathbb{R}^{2}\times \mathbb{CP}^{1}\sim \mathbb{R}_{>0}\times \mathbb{S}^{3}
\end{equation}%
and consequently gravitational line defects $\mathrm{\gamma }_{z}$ spreading
in $\mathbb{R}^{2}$ are dual to particle states on the 3-sphere. Notice also
that for the elliptic case where $\boldsymbol{M}_{4}=\mathbb{R}^{2}\times
\mathbb{T}^{2}$; the 3D boundary has a $\mathbb{T}^{3}$ geometry; then the
dual CS gauge theory lives on a 3-torus. Progress in this direction will be
reported in a future occasion.%
\begin{equation*}
\end{equation*}

\end{document}